%% file: ANA-HION-2018-04-PAPER.tex
\newcommand*{\ATLASLATEXPATH}{}
\documentclass[PAPER, cernpreprint, atlastdraft=true, texlive=2016, UKenglish]{\ATLASLATEXPATH atlasdoc}
 
\PreprintIdNumber{CERN-EP-2019-130}
 
\usepackage{\ATLASLATEXPATH atlaspackage}
\usepackage[backref=false]{\ATLASLATEXPATH atlasbiblatex}
 
\usepackage{\ATLASLATEXPATH atlasphysics}
\usepackage{\ATLASLATEXPATH atlasheavyion}
\usepackage{newtxmath}

\graphicspath{{logos/}{figures/}}
 
\usepackage{ANA-HION-2018-04-PAPER-defs}

\AtlasTitle{
Measurement of flow harmonics correlations with mean transverse momentum in lead--lead and proton--lead collisions at~$\sqrt{s_{\text{NN}}}~=~5.02$~TeV with the ATLAS detector}
 
\AtlasAbstract{
To assess the properties of the quark–gluon plasma formed in ultrarelativistic ion collisions,
the ATLAS experiment at the LHC measures a correlation between the
mean transverse momentum and the flow harmonics.
The analysis uses data samples of lead--lead and proton--lead
collisions obtained at the centre-of-mass energy per nucleon pair of 5.02~TeV,
corresponding to total integrated luminosities of $22~\mu\text{b}^{-1}$
and $28~\text{nb}^{-1}$, respectively.
The measurement is performed using a modified Pearson correlation
coefficient with the charged-particle tracks on an event-by-event basis.
The modified Pearson correlation coefficients for the
2$^{\mathrm{nd}}$-, 3$^{\mathrm{rd}}$-, and 4$^{\mathrm{th}}$-order
flow harmonics are measured in the lead--lead collisions as a function of event centrality quantified as
the number of charged particles or the number of nucleons
participating in the collision.
The measurements are performed for several intervals of the charged-particle transverse momentum.
The correlation coefficients for all studied harmonics exhibit a strong
centrality evolution, which only weakly
depends on the charged-particle momentum range. In the proton--lead
collisions, the modified Pearson correlation coefficient measured for
the 2$^{\mathrm{nd}}$-order flow harmonics shows only weak centrality dependence.
The lead-lead data is qualitatively described by the predictions based on the hydrodynamical model.
 
}
 
\author{The ATLAS Collaboration}
 
\AtlasRefCode{HION-2018-04}

\AtlasHeading{ATLAS Paper}

\AtlasCoverSupportingNote{Internal note}{https://cds.cern.ch/record/2302163}

\AtlasJournalRef{Eur. Phys. J. C 79 (2019) 985}
\AtlasDOI{DOI:10.1140/epjc/s10052-019-7489-6}
 
\AtlasCoverAnalysisTeam{Adam Trzupek, Tomasz Bold, Klaudia Burka}
 
\AtlasCoverEdBoardMember{Alexander, Milov (Weizmann Rehovot) (chair)}
\AtlasCoverEdBoardMember{Krzysztof, Sliwa (Tufts) }
\AtlasCoverEdBoardMember{Radim, Slovak (Prague CU) }

\AtlasCoverEgroupEditors{atlas-HION-2018-04-editors@cern.ch}
 
\AtlasCoverEgroupEdBoard{atlas-HION-2018-04-editorial-board@cern.ch}

\hypersetup{pdftitle={ATLAS document},pdfauthor={The ATLAS Collaboration}}
 
\begin{document}
 
\maketitle

\section{Introduction}
The large azimuthal anisotropy observed for particles produced in
heavy-ion collisions at RHIC~\cite{Back:2004je,Adams:2005dq,Arsene:2004fa,Adcox:2004mh} and the
LHC~\cite{PhysRevLett.105.252302,HION-2011-05,HION-2011-01,CMS-HIN-10-002} is one of the main signatures of the formation of strongly interacting
matter called quark--gluon plasma (QGP).
A standard picture of an ultrarelativistic heavy-ion collision is that the initial, asymmetric `almond' shape of the colliding nuclei's overlap region leads to the formation of pressure gradients in the QGP. These pressure gradients transform the initial shape into an azimuthal anisotropy of the final-state particle distributions through a nearly ideal hydrodynamic evolution and subsequent QGP hadronisation process~\cite{JS}.
The azimuthal anisotropy is customarily decomposed into Fourier
components with the amplitude of the $n^\mathrm{th}$ term denoted by $v_n$ and known as a
flow harmonic~\cite{HION-2016-06}.
Theoretical hydrodynamical models successfully describe observed flow phenomena at low particle transverse momenta~\cite{Ryu}.
The properties of QGP were recently studied with measurements of correlations
between flow harmonics of different order~\cite{QIU2012261,
HION-2014-03,PhysRevC.83.064904,BHALERAO201594,Aad:2013xma} as
well as with analyses of event
shapes~\cite{Adare:2018zkb,Alver:2010rt,Aad:2013xma,ESE,PhysRevC.93.034916}.
It is expected that in lead--lead (\PbPb) collisions the magnitudes of the azimuthal flow harmonics~\cite{HION-2011-05,HION-2011-01} should be correlated with the mean transverse momentum \mpT of the particles on an event-by-event basis~\cite{Heinz:2013th}.
In this paper, that correlation is called the  $v_n$--$\mpT$ correlation.
In proton--lead (\pPb) collisions, the measurements of multi-particle
correlations~\cite{HION-2015-14} show evidence of collective phenomena.
The spectra of identified particles in \pPb\ collisions are consistent with a presence
of the radial flow~\cite{ALICE_pPb_Spectra} while the
nuclear modification factor at high \pt approaches unity~\cite{Acharya:2018qsh}.
Despite intensive studies, the mechanism responsible for the collective behaviour in small collision systems still remains unknown~\cite{JS}.
In \pPb\ collisions the $v_n$--$\mpT$ correlation could provide
constraints on the initial geometry of the particle source,
thereby reducing the overall modelling uncertainty.
According to the hydrodynamical model predictions~\cite{Bozek:2016yoj}, in \pPb\ collisions the $v_n$--$\mpT$
correlation is sensitive to the distribution of energy deposition in the first stage of the collision.
For a larger source a positive $v_2$--$\mpT$ correlation
is expected while for a compact source the negative correlation is obtained.
Simultaneous measurements of $v_n$--$\mpT$ correlations in small and large
systems may help disentangle the role of initial conditions and
subsequent dynamical QGP evolution in final-state particle
distributions.
 
To measure the strength of the $v_n$--$\mpT$ correlation, the Pearson
correlation coefficient (\pcc) $R$~\cite{Bozek:2016yoj}
is used where
\begin{equation}
R = \frac{\covn{n}}{\sqrt{ \var(\vtwo{n}^2)}  \sqrt{ \var(\mpT) }}.
\label{eq:pearson}
\end{equation}
The term $\vtwo{n}^2$ is the square of the $n^\mathrm{th}$-order flow harmonic obtained
using the two-particle correlation method~\cite{JJ-subevent}, \covn{n} is the covariance between $\vtwo{n}^2$ and $\mpT$, and $\var(\vtwo{n}^2)$ and $\var(\mpT)$ are the variances of the $\vtwo{n}^2$ and $\mpT$ distributions, respectively.
Experimentally, however, the finite event-by-event charged-particle track multiplicity results in
an additional broadening of the $\vtwo{n}^2$
and \mpT distributions due to statistical fluctuations.
Thus, the values of the respective
variances are increased, especially for \mpT.
The magnitude of this broadening depends on the choice of kinematic
region and on detector performance, making direct comparisons
between experimental results and with theoretical
calculations difficult.
To overcome this problem, a modified correlation coefficient $\rho$,
less sensitive to the charged-particle multiplicity  than $R$, was suggested in
Ref.~\cite{Bozek:2016yoj}.
To reduce the auto-correlation effects and those due to the finite charged-particle multiplicity in an event, the variances of the $\vtwo{n}^2$
and  \mpT\ distributions are replaced by corresponding {\em
dynamical} variables, which are more sensitive to intrinsic initial-state
fluctuations.
The variance of $\vtwo{n}^2$ is replaced by its dynamical counterpart~\cite{HION-2012-04}
\begin{equation}
\varn{n} = v_n\{2\}^4 - v_n\{4\}^4 = \langle \corr_{n}\{4\}\rangle - \langle \corr_{n}\{2\}\rangle^2,
\label{eq:var}
\end{equation}
where $\corr_{n}\{2\}$ and $\corr_{n}\{4\}$ are the two- and
four-particle correlations~\cite{JJ-subevent} and where angular brackets
denote that they are averaged over events. These correlations are described in detail
in Section~\ref{sec:analysis}.
 
The variance of \mpT is replaced by the dynamical \pT\
fluctuation magnitude~\cite{Adams:2005ka,Abelev:2014ckr} \ck\ defined as
\begin{equation}
\ck =\Big\langle \frac{1}{N_\mathrm{pair}} \sum_i
\sum_{j\ne i}  ({\pT}_{,i} - \langle\mpT\rangle)
({\pT}_{,j}-\langle\mpT\rangle) \Big\rangle
\label{eq:ck}
\end{equation}
 
where $ \langle\mpT\rangle$ is the average $\mpT$ over the all analysed events. The modified \pcc\ $\rho$ is thus defined as
\begin{equation}
\rho = \frac{\covn{n}}
{ \sqrt{ \varn{n} } \sqrt{ \ck \vphantom{(\{\})^x_{xxx}}}}.
\label{eq:rho}
\end{equation}
It was demonstrated in Ref.~\cite{Bozek:2016yoj} that the $\rho$ coefficient
calculated using realistic and finite multiplicities provides a reliable
estimate of the true value of $R$ found in the limit of infinite multiplicity, whereas the coefficient $R$,
calculated using Eq.~\eqref{eq:pearson}  for finite multiplicity underestimates the true value.
 
The ALICE experiment measured~\cite{PhysRevC.93.034916} that the
charged-particle \pT\ spectrum is correlated with the magnitude of
the elliptic flow. It is measured to be harder in collisions with the
higher second flow harmonics and softer in collisions where the
elliptic flow is smaller. The magnitude of spectra modification is
observed to increase with \pT, starting to be significant at around 1~\GeV\ and
reaching a few percent at around 5~\GeV. The modification is found
to be most significant in the mid-central collisions, decreasing in
the most central ones.  The ALICE results suggest that the value of
the correlation coefficient should be significant in mid-central and
central collisions and that its magnitude and centrality dependence
should be sensitive to the scale of intrinsic fluctuations of $v_2$
and \pT. Including particles of higher \pT in the measurement is
expected to result in increased values of the $\rhon{2}$. The $[\pT]$
correlations with $v_2$ in peripheral \PbPb\ collisions, $v_3$ and
$v_4$ in wide centrality range as well as for the $v_2$ in high
multiplicity \pPb\ are unexplored by measurements.
 
This paper reports on the first measurement of the $\rho$ coefficient
with the ATLAS detector in \PbPb\ and \pPb\ collisions at a centre-of-mass
energy per nucleon pair of 5.02~\TeV.
The \PbPb data sample with a total integrated luminosity of 22 $\imub$ was collected
in 2015, and the \pPb\ sample with 28~$\inb$ in 2013.
 
This paper is organised as follows. Section \ref{sec:detector} gives a brief description of the ATLAS detector.
Details of the event selection and charged-particle reconstruction are provided in Section~\ref{sec:eventSel}.
Section~\ref{sec:analysis} describes the analysis procedure for calculating the $\rho$ coefficient. Systematic
uncertainties are described in Section~\ref{sec:Sys} and Appendix A.
Results are presented in Section~\ref{sec:results}, followed by a
summary in Section~\ref{sec:summary}.
 
\section{Experimental setup}
\label{sec:detector}
The ATLAS experiment~\cite{PERF-2007-01} at the LHC is a multipurpose particle detector
with a forward--backward symmetric cylindrical geometry and a near $4\pi$
solid angle coverage. 
The inner detector (ID) covers the pseudorapidity\footnote{ATLAS
uses a right-handed coordinate system with its origin at the
nominal interaction point (IP) in the centre of the detector
and the $z$-axis along the beam pipe. The $x$-axis points from
the IP to the centre of the LHC ring, and the $y$-axis points upward.
Cylindrical coordinates $(r,\phi)$ are used in the transverse plane,
$\phi$ being the azimuthal angle around the $z$-axis.
The pseudorapidity is defined in terms of the polar angle $\theta$
as $\eta=-\ln\tan(\theta/2)$.} range $|\eta| < 2.5$ and
is surrounded by a thin superconducting solenoid providing a \SI{2}{\tesla} axial magnetic field.
The ID consists of silicon pixel, silicon microstrip (SCT), and straw tube tracking detectors.
After the 2013 \pPb run,  an additional pixel silicon layer, the insertable B-layer~\cite{ATLAS-TDR-2010-19,ATLAS-TDR-2010-19,PIX-2018-001}, was installed prior to the 5.02~\TeV\ \PbPb data-taking to attain more precise tracking.
Lead/liquid-argon (LAr) sampling calorimeters provide electromagnetic (EM) energy measurements
with high granularity.
A steel/scintillator tile hadronic calorimeter covers the central pseudorapidity range ($|\eta| < 1.7$).
The endcap and forward regions are instrumented with LAr calorimeters
for EM and hadronic energy measurements up to $|\eta| = 4.9$.
The forward calorimeter (FCal) covers 3.2 $<|\eta|<$ 4.9 and is used for centrality estimation~\cite{HION-2016-06}.
The minimum-bias trigger scintillators (MBTS) are located on each side of the detector at $z = \pm 3.6$~m and detect charged particles with $2.07 < |\eta| < 3.86$.
The zero-degree calorimeter (ZDC), located in the LHC tunnel and covering
$|\eta|>8.3$, is used for triggering on collision events and pile-up event rejection.
It is calibrated to resolve an individual neutron originating from the collision spectators.

A two-level trigger system selects events~\cite{TRIG-2011-02,Aaboud:2016leb}. The level-1 trigger is implemented in hardware and preselects
up to $10^5$ events per second for further decisions by the high-level trigger (HLT).
The software-based HLT tuned for \PbPb\ collision data selects up to 1000 events per second for recording.
This analysis primarily uses charged-particle tracks in the ID,
but information from the central calorimeters and the ZDC is also used for triggering, event selection, and analysis.
 
 
\section{Event and track selection}
\label{sec:eventSel}
The \PbPb\ data in this analysis were selected using two mutually exclusive minimum-bias triggers.
Events with semi-central and central collisions were selected if the scalar sum of transverse energy in the entire ATLAS calorimeter system exceeded 50~\GeV.
Peripheral events, i.e.\ those with large impact parameter of the colliding Pb nuclei, fail the 50~\GeV\ selection and were instead selected by requiring a deposition in the ZDC corresponding to at least one neutron and by requiring at least one track reconstructed in the HLT.
Data in this analysis are required to come from periods when the entire detector was functioning normally.
The events are required to have a reconstructed vertex within \SI{100}{mm} of the nominal interaction point.
The contribution from events containing more than one inelastic interaction (pile-up)
is studied by exploiting correlations
between the transverse energy measured in the FCal (\fcalsum) with
the estimated number of neutrons in the ZDC, and  with the number of tracks associated with a primary vertex~\cite{HION-2012-04,Aaboud:2019sma}.
The distribution of \fcalsum and the distribution of the number of neutrons in events with more than one collision are broader than the corresponding distributions in events with only one collision.
Pile-up events are suppressed by rejecting events with abnormally large values of either \fcalsum\ or the number of neutrons in the ZDC compared with the charged-particle multiplicity in the event.  Approximately 0.2\% of the events are rejected
by these requirements.
 
The \pPb data  in this analysis were selected using minimum-bias triggers and high-multiplicity triggers (HMT).
The minimum-bias trigger required signals in both sides of the MBTS system with a timing difference of less than 10~ns to eliminate non-collision backgrounds.
The HMT required the total transverse energy in the calorimeter at level-one and the number of ID track candidates reconstructed in the HLT to be above predefined thresholds.
Six combinations of thresholds were used to optimise data-taking during periods with different luminosities.
Samples of events collected by these triggers are combined by applying event weights to reproduce the charged-particle multiplicity distribution of the minimum-bias trigger.
Further details of the data selection are given in Refs.~\cite{HION-2015-14,HION-2013-04}.
The average pile-up probability in the \pPb dataset is approximately 3\% but can be significantly larger in high-multiplicity events.
Events with more than one reconstructed vertex are removed from the sample.
Similarly to the Pb+Pb dataset, to remove events where the two interaction vertices
are too close to resolve as independent ones,
the ZDC signal on the Pb fragmentation side is used.
The distribution of the number of neutrons, which is broader in events with pile-up than that for the events without pile-up is exploited for that purpose~\cite{HION-2013-04}.
The fraction of rejected events varies with the event activity and reaches a maximum of 10\% for events with the highest multiplicities.
 
The analysis for both collision systems is performed in narrow bins of event activity defined by the charged-particle
multiplicity \Nch (described in  \Sect{{\ref{sec:analysis}}}), which estimates the collision centrality.
In addition, the \PbPb\ results are presented as a function of collision centrality expressed by the average number of nucleons participating in the collision, \Npart, to allow comparison with theoretical predictions~\cite{Bozek:2016kpf}. The centrality is estimated from the \fcalsum distribution~\cite{HION-2011-05,HION-2016-06} using the Glauber model~\cite{GlauberMC}.
The number of events passing the selection requirements is $1.3\times 10^8$ for \PbPb within the 0--80\% centrality interval. For the \pPb system, about $0.64 \times 10^8$ events enter the analysis.
 
The charged-particle tracks reconstructed in the ID are required
to satisfy selection criteria in order to suppress the contribution
of incorrectly reconstructed tracks and secondary products of particle decays.
The selection criteria include the requirement that the number of hits in the pixel and
SCT detectors should be greater than two and eight, respectively, for the \PbPb data and greater than one and six for the \pPb data.
The track impact parameters relative to the collision vertex in the transverse direction, $|d_0|$, and longitudinal direction, $|z_0\sin\theta|$,
are required to be less than 1~mm for tracks in the \PbPb data sample and less than 1.5~mm in the \pPb sample.
In addition, in \pPb collisions, the track impact parameter significances must satisfy $|d_0/\sigma_{d_0} | <$ 3 and $|z_0 \sin\theta /\sigma_{z} | <$ 3, where $\sigma_{d_0}$  and $\sigma_{z}$ are the uncertainties in $d_0$ and $z_0 \sin\theta$ determined from the covariance matrix of the track fit.
The different selection criteria for \PbPb and \pPb optimise the performance of the track reconstruction in differing running conditions.
 
Corrections needed due to track reconstruction effects are evaluated using 4$\times10^6$ \PbPb and $10^7$ \pPb minimum-bias Monte Carlo (MC) events generated by the HIJING v1.38b~\cite{PhysRevD.44.3501} event generator. After the generation, an azimuthal flow is implemented using the afterburner technique~\cite{Jia:2013tja}, and the \pT\ spectrum is reweighted to match the data. Generated events were simulated in the detector by the \textsc{Geant 4}-based~\cite{Agostinelli:2002hh} ATLAS detector simulation programs~\cite{SOFT-2010-01} and reconstructed using the same procedures and detector conditions as the data.
Track reconstruction corrections are applied to each selected track using weights to account for the tracking efficiency $\epsilon$ and the fake-track fraction $f$. The efficiency is defined as the fraction of primary MC charged particles that are matched to reconstructed tracks, and $f$ is the fraction of tracks that are not matched to primary MC particles or are produced from random combinations of hits in the ID. A similar analysis procedure is described in Refs.~\cite{Aad:2013xma,HION-2016-06}. The fake-track fraction and tracking efficiency are determined as functions of the track \pT and $\eta$ and of the track multiplicity in the event. Tracks included in the analysis are weighted with the factor $(1-f)/\epsilon$.
An additional multiplicative weight evaluated from data is applied to the data to correct for detector non-uniformity in the azimuthal angle. These weights are obtained by requiring the tracks to be distributed uniformly in azimuth in all pseudorapidity slices of width 0.1.

In the \PbPb data, the contribution of fake tracks is largest in central collisions at the lowest analysed track \pT\ of 0.5~\GeV\ and at the largest $|\eta|$, reaching up to 20\%. The fake-track rate is below 1\% for tracks with \pT\ above 2~\GeV\ and $|\eta|<1.5$. The tracking efficiency depends weakly on centrality, and in the most central events it is about 3\% less than in more peripheral events. The efficiency increases with the track \pT\ from about 50\% at the lowest analysed \pT\ to 70\% above 2~\GeV. It is highest at mid-rapidity and drops by about 15\% for $|\eta|>1$.
For \pPb collisions, with \pT increasing from 0.3 to 1~\GeV\ the efficiency increases from about 75\% (60\%) to 83\% (70\%) at $\eta \approx$ 0 ($|\eta| > 2$).  The \pPb tracking efficiency is independent of the event's multiplicity for $\Nch \ge 10$, i.e.\ in the multiplicity range used in the analysis. The fake rate in \pPb collisions is very low,  below 1\% (3\%) at  $\eta \approx$ 0 ($|\eta| > 2$).
 
\section{Correlation coefficient $\rho$}
\label{sec:analysis}
In each event, charged-particle tracks are grouped into three
regions of subevents based on their pseudorapidity:
region A with $-2.5 <\eta<-0.75$, central region B with $|\eta|<0.5$
and region C with $0.75  < \eta < 2.5$.
The $v_{n}^2$ for the $n$ = 2--4 harmonics are calculated by correlating charged-particle tracks from subevents A and C,
which are separated in pseudorapidity to suppress non-flow contributions.
Tracks in central region B are used to obtain the mean value of the charged-particle transverse momentum in the event, $[\pT]$, defined as
\begin{equation*}
\mpT = \frac{1}{\sum_{b}{w_b}} \sum_{b } w_b p_{\mathrm{T}b}
\end{equation*}
where the summation is over tracks in region B,
labelled by index $b$. The variable \ck\ (Eq.~\eqref{eq:ck}) is also calculated using
tracks from region B.
Here, and in following formulas, the weights $w$ include the fake-track fraction, efficiency, and azimuthal
non-uniformity corrections, as discussed in Section~\ref{sec:eventSel}.
 
The covariance term from the numerator of Eq.~\eqref{eq:rho} is defined as
\begin{equation}
\cov (\vnpT{n}) = \mathrm{Re}\Big(\Big\langle \frac{1}{\sum_{a,c}{w_a w_c}} \sum_{a, c } w_a w_c \mathrm{e}^{\mathrm{i}n\phi_a - \mathrm{i}n\phi_c}  (\mpT - \mmpT) \Big\rangle\Big),
\end{equation}
where $\phi$ is the azimuthal angle
and indices $a$ and $c$ span the tracks in regions A and C, respectively.
 
The two- and four-particle correlations used to define the dynamical
variance in Eq.~\eqref{eq:var}, which enters the denominator of
Eq.~\eqref{eq:rho}, are calculated as in Ref.~\cite{JJ-subevent}
\begin{equation}
\langle \corr_{n}\{2\} \rangle = \mathrm{Re} \Big(\Big\langle \frac{1}{\sum_{a,c}{w_a
w_c}} \sum_{a, c } w_a w_c \mathrm{e}^{\mathrm{i}n\phi_a - \mathrm{i}n\phi_c} \Big\rangle\Big) =
\mathrm{Re} \big(\langle q_{n,a} q_{n,c}^{*} \rangle\big)
\label{eq:dyn_var}
\end{equation}
where the $q_a$ and $q_c$ are the complex flow vectors of subevent A and
subevent C, respectively, and the asterisk denotes the complex conjugate. The flow vectors are
\begin{equation*}
q_{n,a}=\frac{1}{\sum_{a}{w_a}} \sum_{a}{w_a \mathrm{e}^{\mathrm{i}n\phi_a}}\hspace{1cm} \mathrm{and} \hspace{1cm}
q_{n,c}=\frac{1}{\sum_{c}{w_c}} \sum_{c}{w_c \mathrm{e}^{\mathrm{i}n\phi_c}}.
\end{equation*}
The four-particle correlation is obtained from the expression
\begin{equation}
\langle \corr_{n}\{4\} \rangle = \mathrm{Re}\Big(\Big\langle
\frac{(Q_{n,a}^2-Q_{2n,a})(Q_{n,c}^2-Q_{2n,c})^{*}}{S_a S_c}
\Big\rangle \Big) ,
\label{eq:cor4}
\end{equation}
where for subevent A
\begin{equation*}
Q_{n,a}=\sum_{a}{w_a \mathrm{e}^{\mathrm{i}n\phi_a}}, \hspace{1cm}
Q_{2n,a}= \sum_{a}{w_a^2 \mathrm{e}^{\mathrm{i}2n\phi_a}}, \hspace{1cm} S_a= \Big(\sum_{a}{w_a}\Big)^2- \sum_{a}{w_a}^2,
\end{equation*}
and similarly for subevent C. Equation~\eqref{eq:cor4} represents the sum $\sum{\mathrm{e}^{\mathrm{i}n(\phi_1^a+\phi_2^a - \phi_3^c - \phi_4^c)}}$ over all particles from subevents A and C normalised by the number of quadruplets without auto-correlations in each subevent.
 
The second factor in the denominator of Eq.~\eqref{eq:rho}, the mean
\pT\ fluctuation in the event class \ck, is defined by Eq.~\eqref{eq:ck} and in this analysis it is calculated as
\begin{equation*}
\ck = \Big\langle \frac{1}{(\sum_b w_b)^2 - \sum_b w_b^2} \sum_b
\sum_{ b'\ne b} w_b  (p_{\mathrm{T},b} - \mmpT) w_{b'}
(p_{\mathrm{T},b'}-\mmpT) \Big\rangle.
\end{equation*}
The summation indices  $b$ and $b'$  run over all charged particles in region B.
 
The correlation coefficient expressed by Eq.~\eqref{eq:rho} is evaluated for the range  0.5~$<\pT<$~2~\GeV\ in  \PbPb collisions and 0.3~$<\pT<$~2~\GeV\ in \pPb collisions. These intervals, called `main', contain a large number of soft particles and constitute the
main result of the analysis which can be compared with hydrodynamical models.
For each system, two additional \pT\ ranges are considered:
0.5~$<\pT<$~5~\GeV\ and 1~$<\pT<$~2~\GeV\ in the analysis of
\PbPb collisions, and 0.3~$<\pT<$~5~\GeV\ and 0.5~$<\pT<$~2~\GeV\
in \pPb collisions. These ranges facilitate the study of the sensitivity of $\rhon{n}$ to
the high \pt part of the particle spectrum 
and to the lower charged-particle multiplicity from the higher minimum \pT value.
The charged-particle \pT\ range 0.5~$<\pT<$~2~\GeV\
is common to both systems and can be used to compare
the $\rhon{2}$ results from \PbPb and \pPb collisions.
 
The quantities of interest, i.e.\ $\covn{n} $, $\varn{n}$, \ck, and \rhon{n},
are determined in bins of reconstructed track multiplicity \mulac\ measured in
the combination of regions A and C.
This is done to avoid a negative correlation between
the multiplicity in subevents A+C and B that
occurs if the analysis is binned in multiplicity in the entire ID.
Narrow \mulac bins are also chosen due to the sensitivity to
multiplicity fluctuations of the multi-particle correlations that are used to obtain the
\varn{n}~\cite{HION-2012-04}.
The events are grouped in fine bins with a width of ten in \mulac for 0.5~$<
\pT <$ 5~\GeV\ in the \PbPb\ analysis and 0.3~$< \pT <$ 5~\GeV\ in the \pPb\ analysis.
It was cross-checked that the variables of interest obtained with a
finer binning in \mulac are consistent with the
measurement with the nominal binning.
 
To enable comparisons with the theoretical predictions and with future
experimental results, measurements obtained in \mulac\ are presented as a
function of the ATLAS ID multiplicity \Nch of $0.5 <  \pT < 5~\GeV$
and $|\eta|<2.5$.
They are projected from the \mulac\ values
taking into account tracking efficiency and fake-track production as described in the previous section.
A similar analysis procedure is described in Ref.~\cite{HION-2015-14}.
For the \Npart\ dependencies in the \PbPb\ system, the results measured in \mulac\
multiplicity intervals are averaged, with weights equal to the probabilities to find any given \mulac\ value in the centrality intervals.

The formulation of the modified \pcc\ $\rhon{n}$ requires
that there should be at least two tracks in each region (A, B, and C).
Further, \varn{n} calculated according to Eq.~\eqref{eq:dyn_var} can be
negative  at low multiplicities due to
statistical fluctuations,  which renders Eq.~\eqref{eq:rho} invalid because of
the $\surd\varn{n}$ term.
For each \mulac\ bin, \pT interval, and harmonic, a criterion is applied that \varn{n} needs to be positive at a level of at least one standard deviation of its statistical uncertainty. Results presented as a function of \Nch are produced only for those \mulac\ intervals.
For the \Npart dependencies in the Pb+Pb system, it is additionally required for each centrality interval that the fraction of rejected events due to this criterion does not exceed 1\%.
 
\section{Systematic uncertainties}
\label{sec:Sys}
The systematic uncertainty is estimated by varying
individual aspects of the analysis.
The systematic uncertainties for the main \pT\ interval are discussed
for each collision system.
Systematic uncertainties for the other \pT\ intervals behave
consistently with the ones for the main \pT\ interval.
Since the modified \pcc\ $\rhon{n}$ is a ratio of quantities which are calculated using tracks,
many variations largely cancel out and the resulting systematic uncertainties are small.
To suppress the statistical fluctuations and to get more robust estimation of systematic uncertainties, they are averaged over several, wide ranges of the charged-particle multiplicity.
For each uncertainty source and for each measurement point, the maximum variation from the baseline measurement is used.
The total resulting uncertainty is the sum of the individual contributions combined in quadrature.
The following sources of systematic uncertainties are considered.
 
{\textbf{Track selection.}}
The tracking performance has a relatively small impact on \vtwo{n}, but it directly affects the \mpT\ and \ck via the admixture of the fake tracks, especially at low \pT.
To assess the impact on $\rhon{n}$, the measurement is repeated with
tracks selected with looser and tighter track quality criteria, thus
increasing and decreasing the fake-track rate, respectively.
The weights used in the evaluation of measured quantities take the
modified selection into account.
The loose track selection in the \PbPb\ analysis relaxes requirements on the number of
pixel and SCT hits to at least one and six, respectively.
Additionally, the requirements on the transverse and longitudinal impact parameters of the track are relaxed to 1.5~mm.
The tighter selection in the \PbPb\ analysis tightens the requirement on the transverse and longitudinal impact parameters of the track to 0.5~mm.
For the \pPb\ analysis, the loose selection relaxes the requirements on the transverse and longitudinal impact
parameters of the track to 2~mm and on the impact parameter
significances to less than 4.
In the tight selection, the impact parameter values and their significances must be less than 1~mm and 2, respectively.
For each of the two track selections the absolute difference is calculated with respect to the baseline measurement: $|\rhon{n}^\mathrm{base}- \rhon{n}^\mathrm{loose}|$ or $|\rhon{n}^\mathrm{base}- \rhon{n}^\mathrm{tight}|$. The largest difference is taken as a systematic uncertainty.
 
{\textbf{Detector material.}}
Since the tracks that are used in the calculation of $\rhon{n}$ are weighted by the inverse of the tracking efficiency, a bias in its estimation due to inaccurate modelling of the  material in the detector may change the balance between low- and high-\pT tracks in the sums.
Based on simulations, the estimated uncertainty in the detector description is obtained~\cite{Aaboud:2017pjd,STDM-2015-02}.
The resulting \pT- and $\eta$-dependent uncertainties in the track efficiency of up to 4\% are used to determine the systematic uncertainty.
 
{\textbf{Tracking azimuthal uniformity.}}
In this analysis, the weighting factors $w$ correct for
any non-uniformity in the azimuthal angle distribution of
reconstructed tracks.
The weights are obtained from the data by requiring azimuthal
uniformity over the two-dimensional distribution of reconstructed tracks
in the $\eta$--$\phi$ plane.
The effect of that correction on the result is  conservatively estimated by comparing the baseline measurement and the measurement obtained without applying this weight.
The uncertainty is small, and it envelopes potential effects of imperfections in the weighting factors determination, including their dependence on the transverse momentum, collision centrality, run-by-run differences, on dead module maps or the vertex position.
 
\renewcommand{\arraystretch}{0.8}
\begin{figure}[!htb]
\centering
\begin{tabular}{cc}
\includegraphics[width=0.45\textwidth]{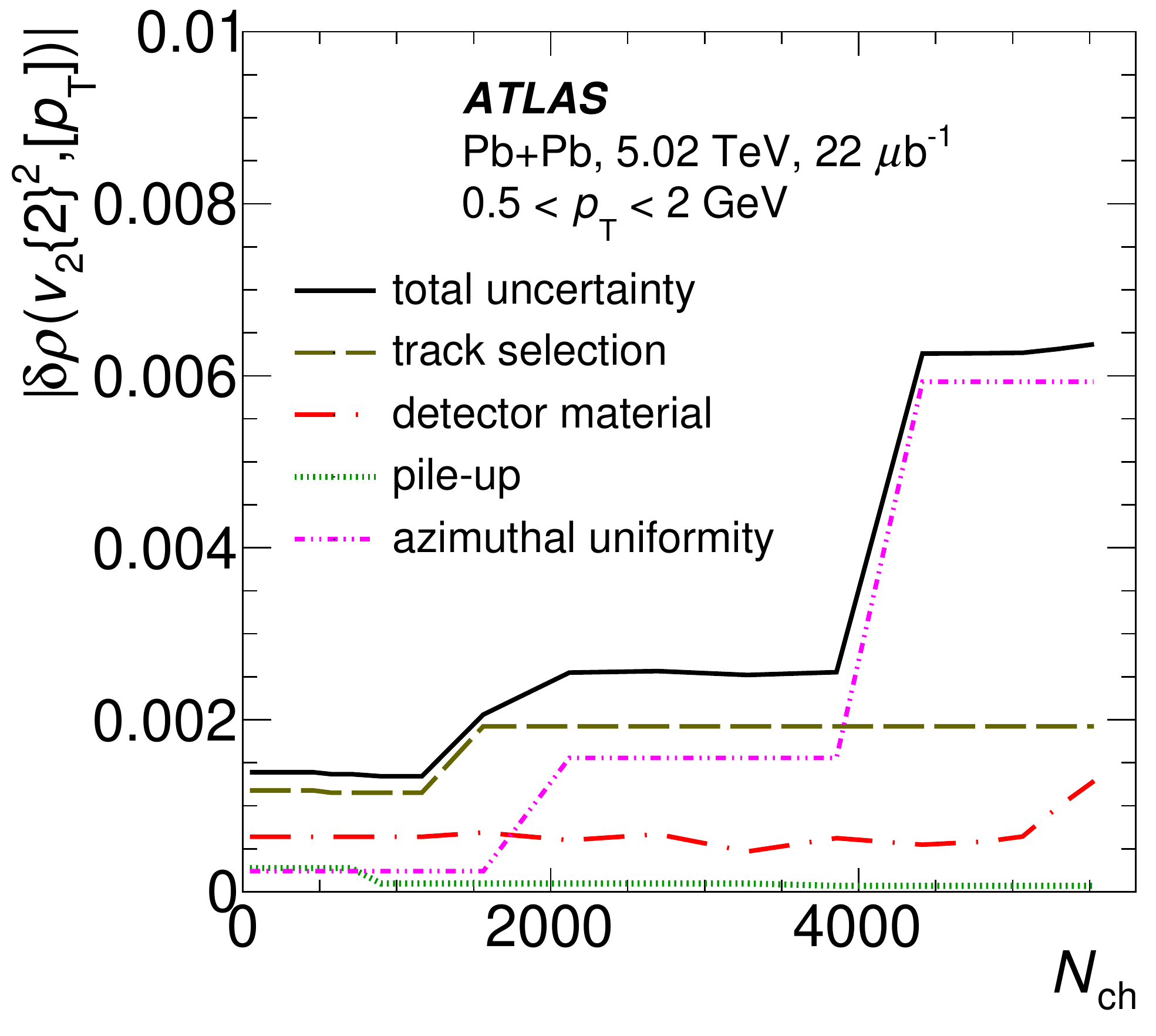} &
\includegraphics[width=0.45\textwidth]{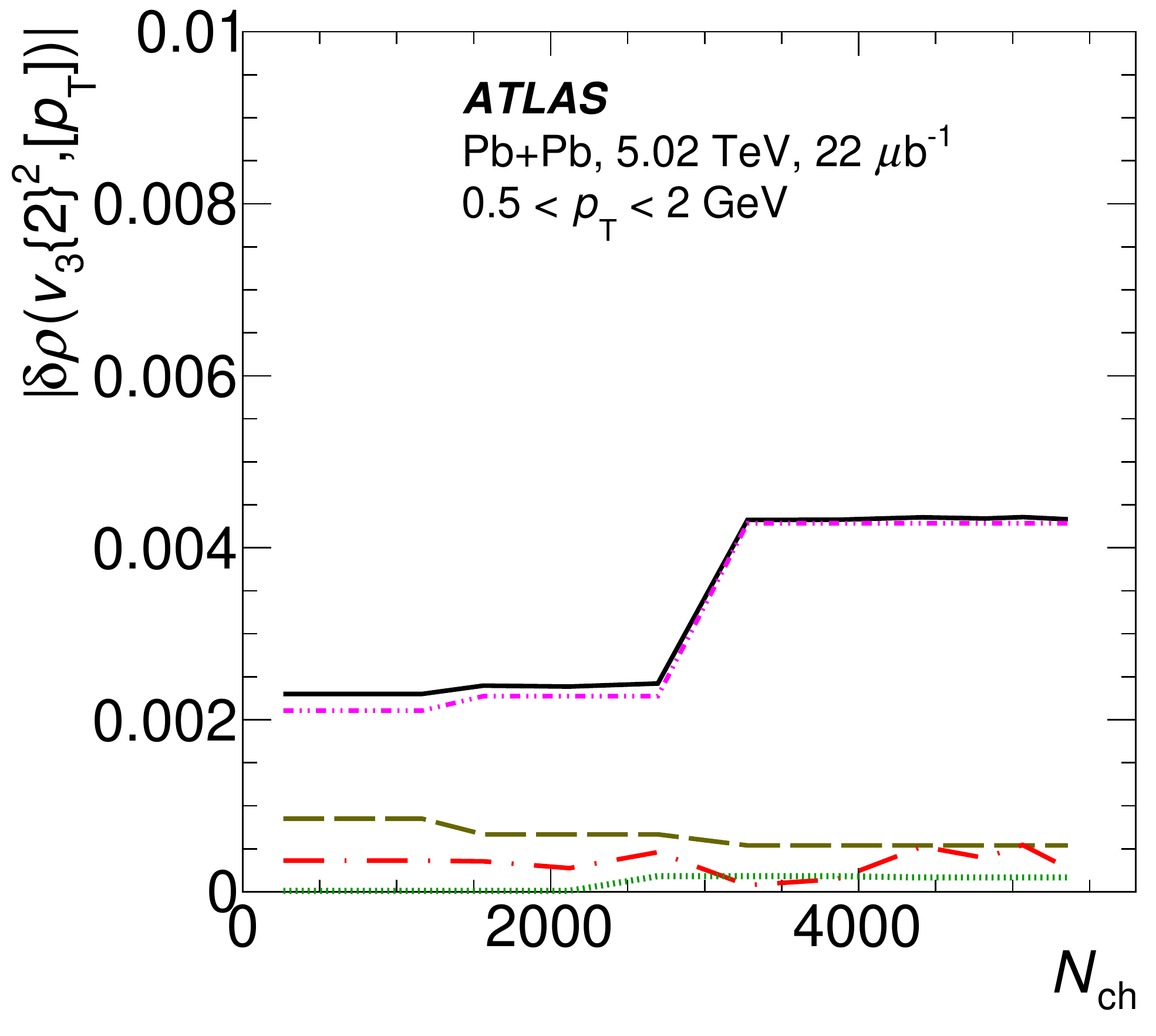}\\
\hspace{1.5cm}(a) &     \hspace{1.5cm}(b) \\
 
\includegraphics[width=0.45\textwidth]{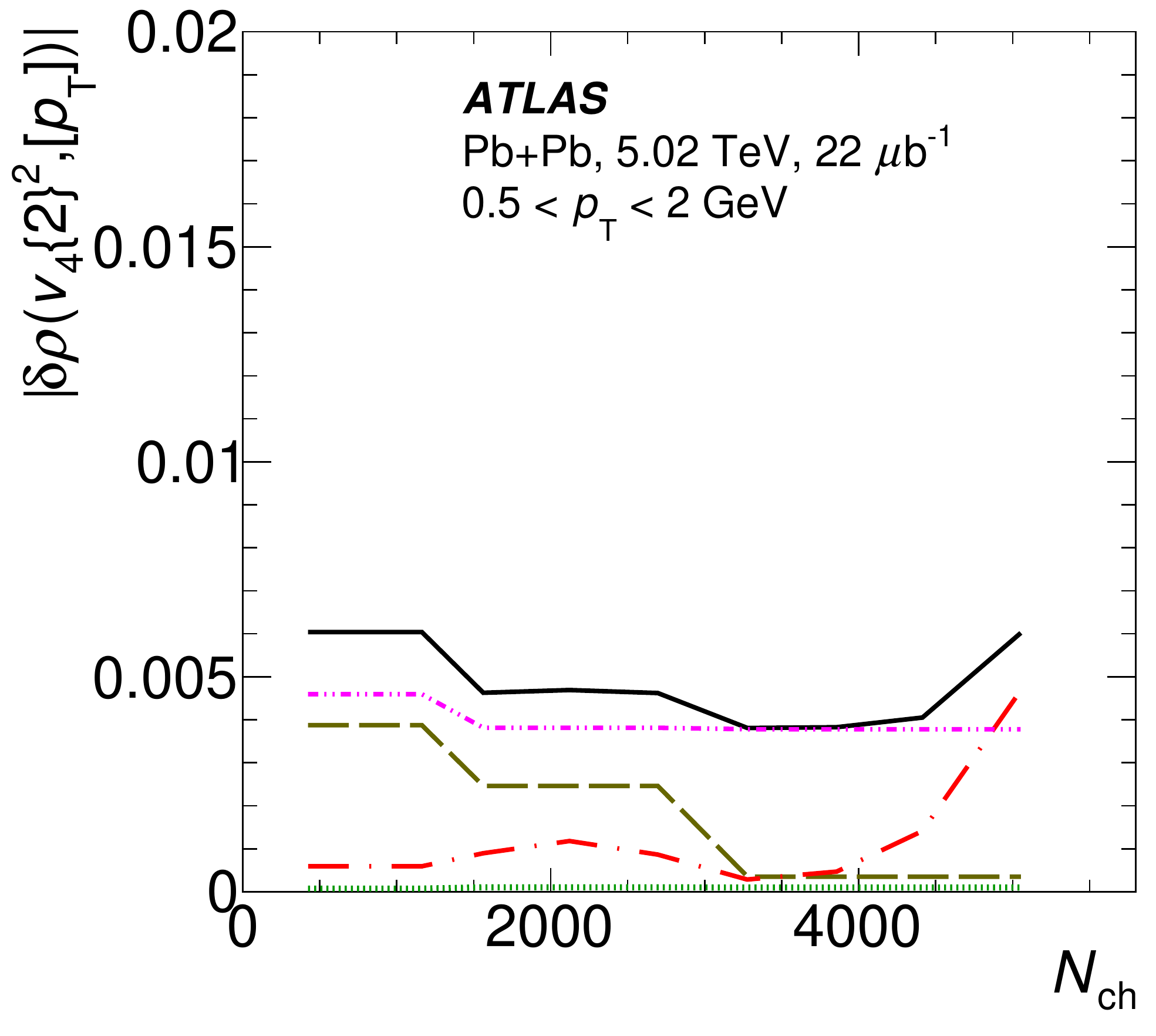} &
\includegraphics[width=0.45\textwidth]{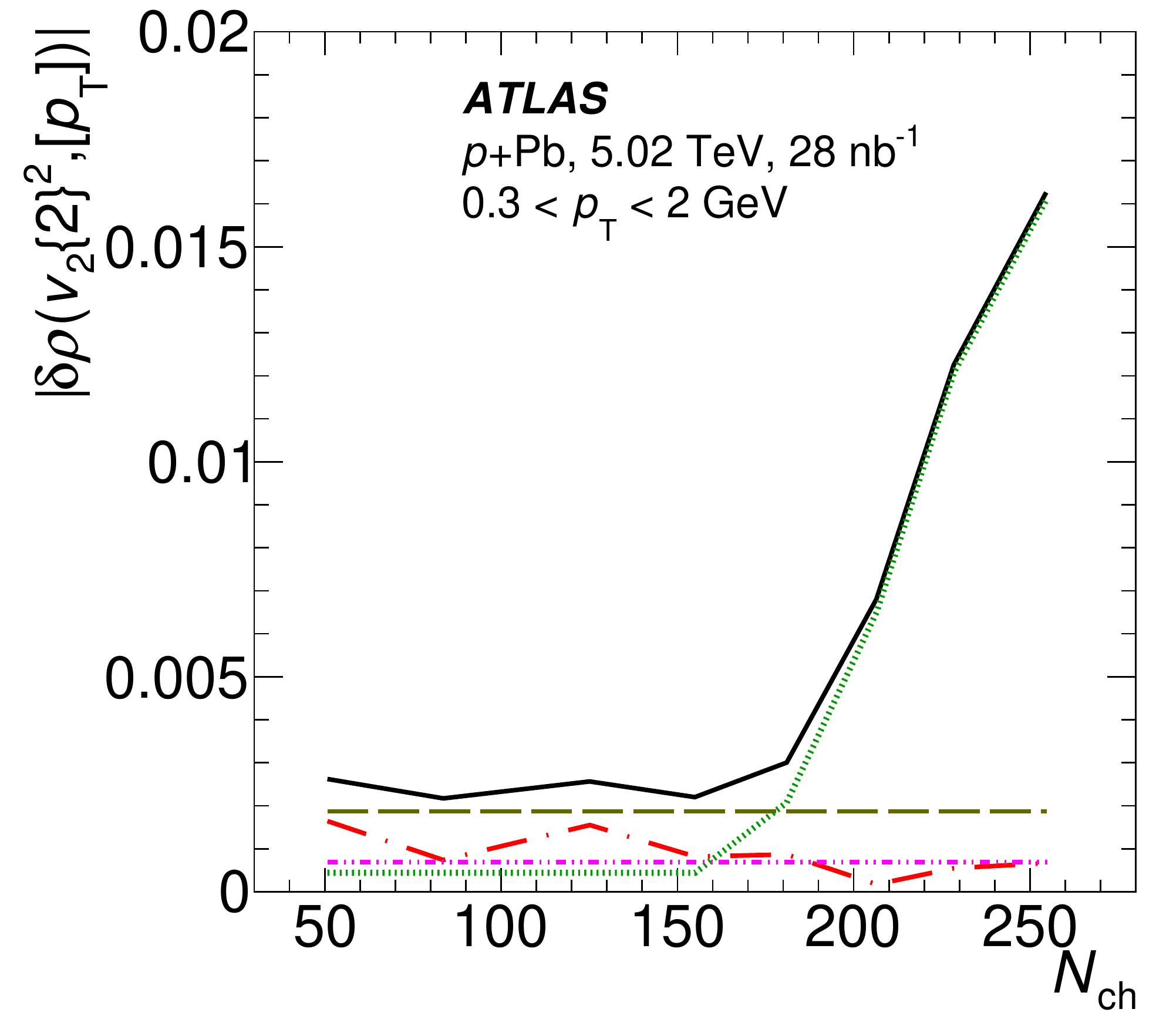}    \\
\hspace{1.5cm}(c) &     \hspace{1.5cm}(d) \\
\end{tabular}
\caption{ The systematic uncertainty of \rhon{n}\ as a
function of \Nch\ measured with tracks from main \pT\ intervals
for each collision system for the (a) second, (b) third, and (c) fourth harmonics
in \PbPb\ collisions,
and for (d) \rhon{2} in \pPb\ collisions. The total uncertainty is also shown.}
\label{fig:sys}
\end{figure}
 
{\textbf{Residual pile-up events.}}
The selection criteria discussed in Section~\ref{sec:eventSel} suppress
the fraction of  pile-up events accepted for analysis to almost zero in central
\PbPb\ collisions.
To estimate the systematic uncertainty related to pile-up, the measurement is
conservatively repeated without this event rejection, resulting
in at most  a 1\% difference in the most central \PbPb\ events for the
\rhon{2} coefficient.
The \pPb data were taken with higher pile-up than the \PbPb\ data.
To estimate the impact of contamination by residual pile-up events, \pPb results were obtained with only the vertex criteria applied. The variation covers the estimated residual pile-up fraction in events of the highest track multiplicity~\cite{HION-2013-04}.
 
{\textbf{Centrality selection.}}
The minimum-bias trigger is fully efficient for the 0--85\% centrality interval. However,
the total fraction of inelastic \PbPb\ events selected is known only to 1\% accuracy
due to trigger inefficiency and possible sample contamination in more peripheral interactions.
The centrality is estimated using the \fcalsum distribution~\cite{HION-2011-05,HION-2016-06} and the Glauber model~\cite{GlauberMC} to obtain the mapping from the observed \fcalsum to the number of nucleons participating in the collision, \Npart.  The modified \pcc\ uncertainty is evaluated by repeating the analysis with the altered centrality selections on the \fcalsum distribution, which results in $\pm 1$\% uncertainty in the total fraction of inelastic \PbPb\ events.
The centrality selection contributes mainly to uncertainties for peripheral collisions.
 
Figure~\ref{fig:sys} shows the magnitude of the systematic uncertainties
$\deltaup\rhon{n}$ for $n$ = 2--4 in \PbPb collisions as a function of \Nch.
In \PbPb collisions, the systematic uncertainty of the measured correlation coefficients across different order harmonics and centralities is not dominated by a single source.
One of the largest uncertainties comes from restoring the azimuthal uniformity, and dominates for the second order harmonic in the most central collisions and for the third and fourth order harmonics almost over the full centrality range.
A sizeable contribution to the uncertainty for all three harmonics is due to the track selection.  The impact of the detector material is rather small except for a significant contribution for the forth order harmonic in the most central events. The residual pile-up in \PbPb collisions gives a negligible contribution.
Figure~\ref{fig:sys}(d) shows systematic uncertainties for $\rhon{2}$ coefficients in \pPb collisions for the main interval of  0.3 $< \pT <$ 2~\GeV\ as a function of event activity.
In \pPb interactions the largest uncertainty in the most active collisions ($\Nch> 150$) originates from pile-up.
The track selection is a source of sizeable uncertainty for this collision system, while the azimuthal uniformity correction procedure and the detector material have a small impact.
 
Details on the contributions to systematic uncertainties from different sources of \ck, \varn{n}\ and \covn{n}\  are included in the Appendix.

 
\section{Results}
\label{sec:results}
\subsection{The constituents of the modified \pcc }

\begin{figure}[!htb]
\centering
\begin{tabular}{cc}
 
\includegraphics[height=0.4\textwidth]{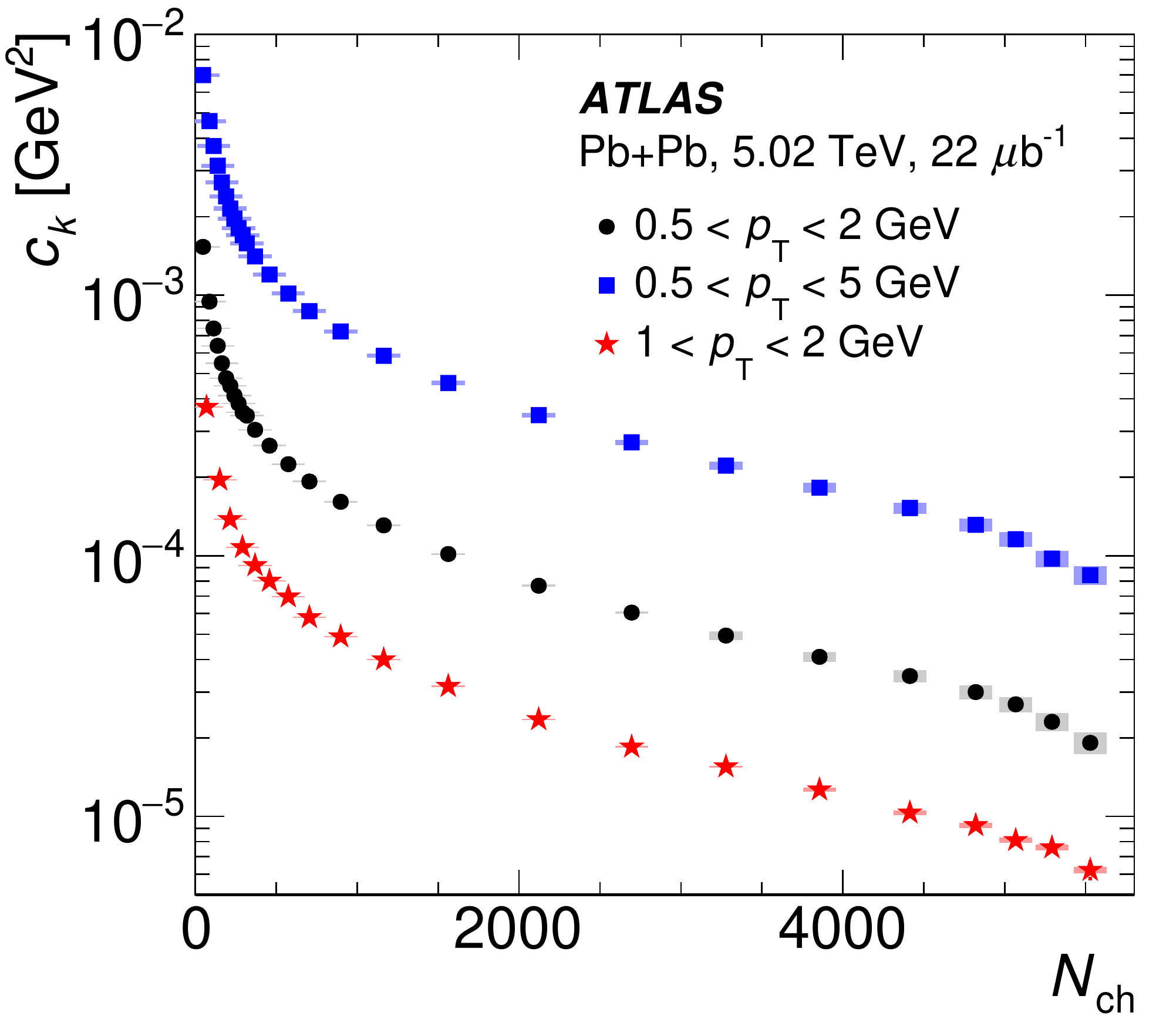} &
\includegraphics[height=0.4\textwidth]{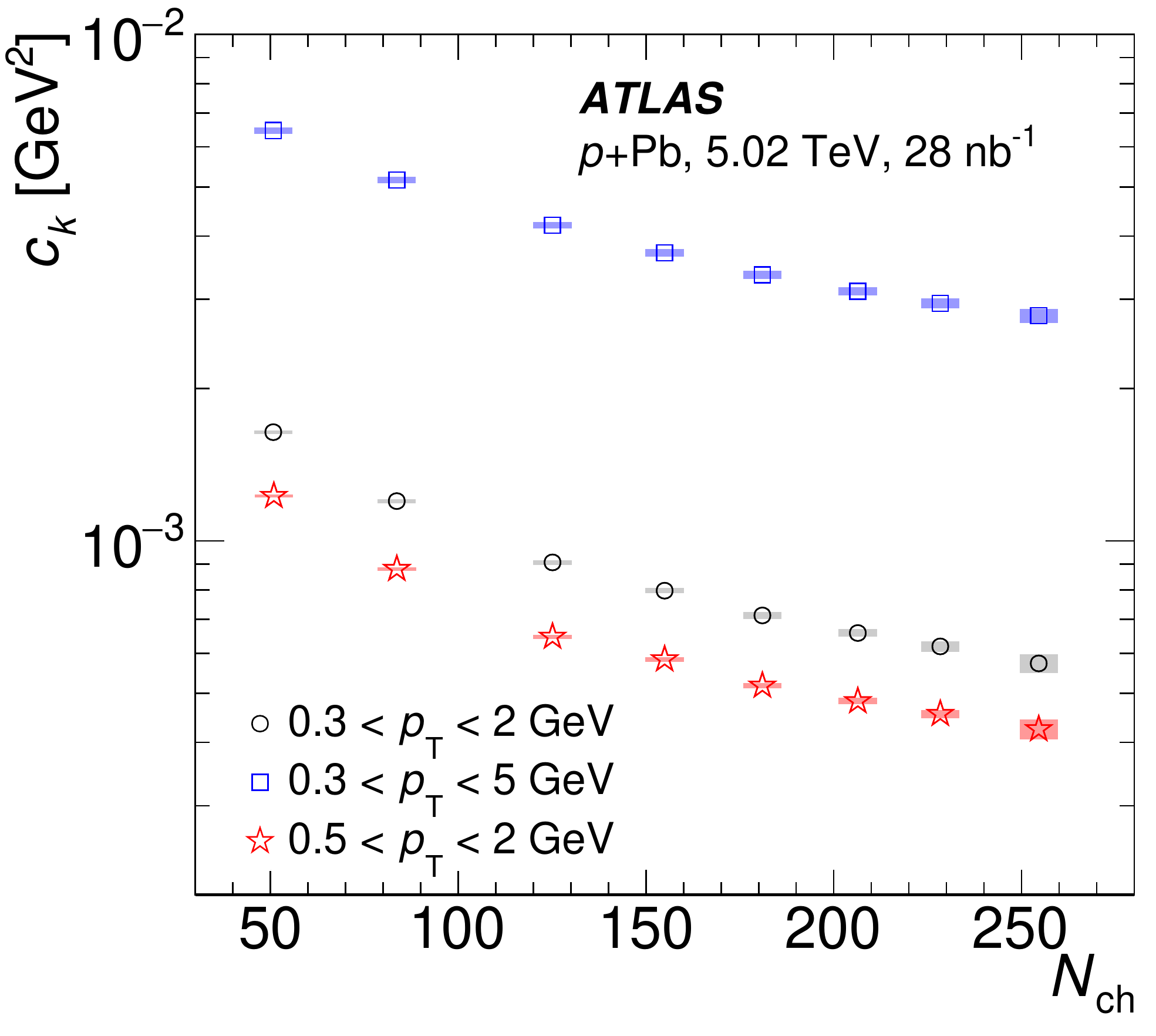} \\
\hspace{1.5cm}(a) &     \hspace{1.5cm}(b) \\
\end{tabular}
 
\caption{ The variable \ck\ for three \pT\ ranges as a function of the charged-particle
multiplicity \Nch\ of (a) \PbPb  and (b) \pPb collisions.
The statistical and systematic uncertainties are shown as vertical error bars (smaller than symbols) and boxes, respectively. }
\label{fig:ck}
\end{figure}

The constituents of the modified \pcc, \ck\ , $\varn{n}$ and  $\covn{n}$ are combined, using Eq.~\eqref{eq:rho}, to obtain $\rho$.
Figure~\ref{fig:ck} shows the dynamical \pT\ fluctuation coefficient
\ck\ as a function of charged-particle multiplicity in \PbPb\ and \pPb\ collision
systems for tracks in three different \pT\ intervals.
A strong decrease of \ck\ with increasing \Nch\ is observed in all measured results.
A similar decrease was seen for \ck\ in Au+Au and \PbPb\ data at lower
centre-of-mass energies~\cite{Adams:2005ka,Abelev:2014ckr}, evaluated for lower \pT range, $0.15<\pT < 2$~\GeV, not accessible with the ATLAS detector.
For the same \Nch, the \ck\ values differ by an order of
magnitude for different \pT\ ranges of tracks used in the analysis.
For the intervals with the same lower \pT\ limit, the \ck\ values are higher for the interval with the larger upper \pT\ limit.
 
\begin{figure}[!htb]
\centering
\begin{tabular}{cc}
 
\includegraphics[width=0.45\textwidth]{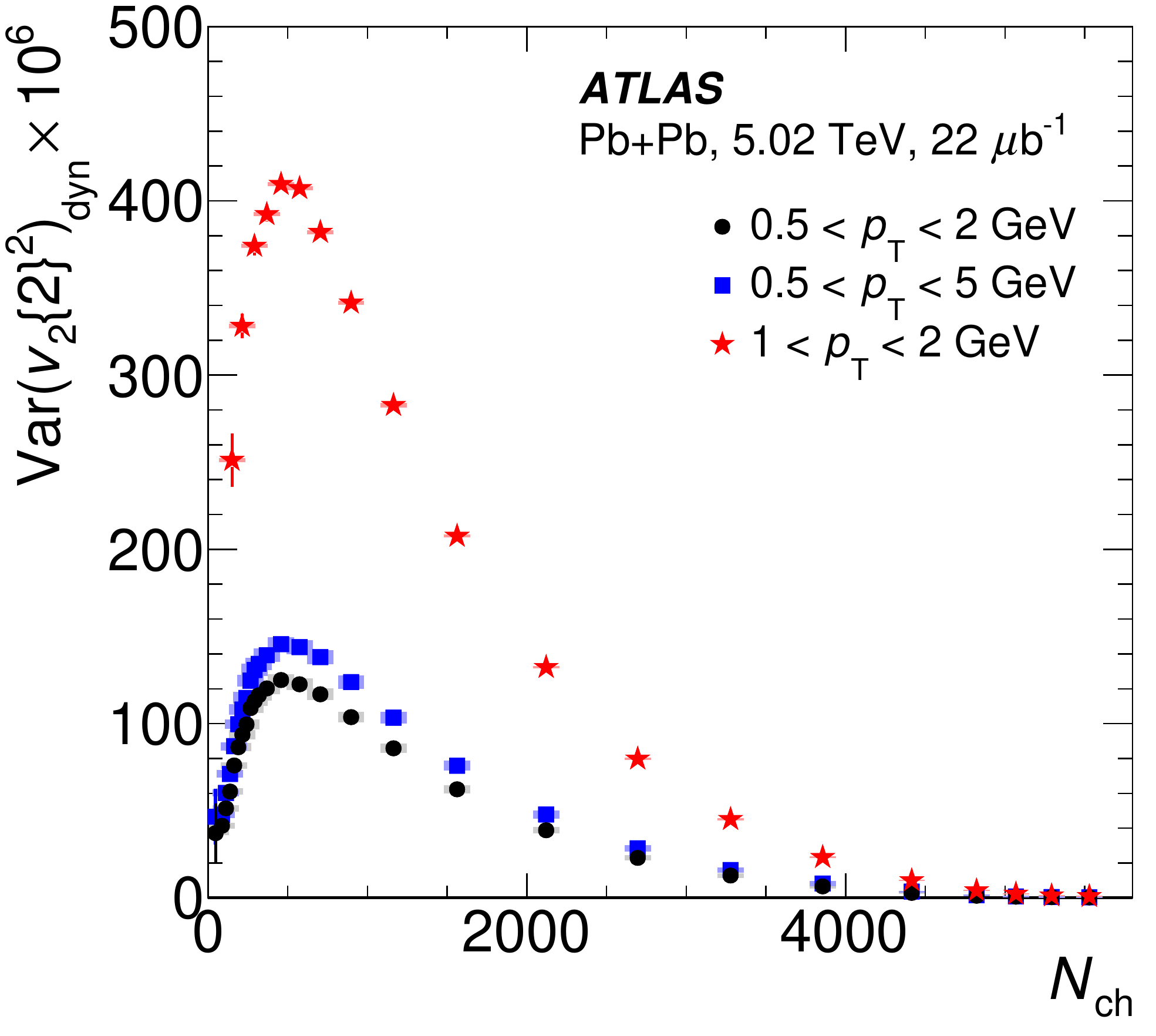} &
\includegraphics[width=0.45\textwidth]{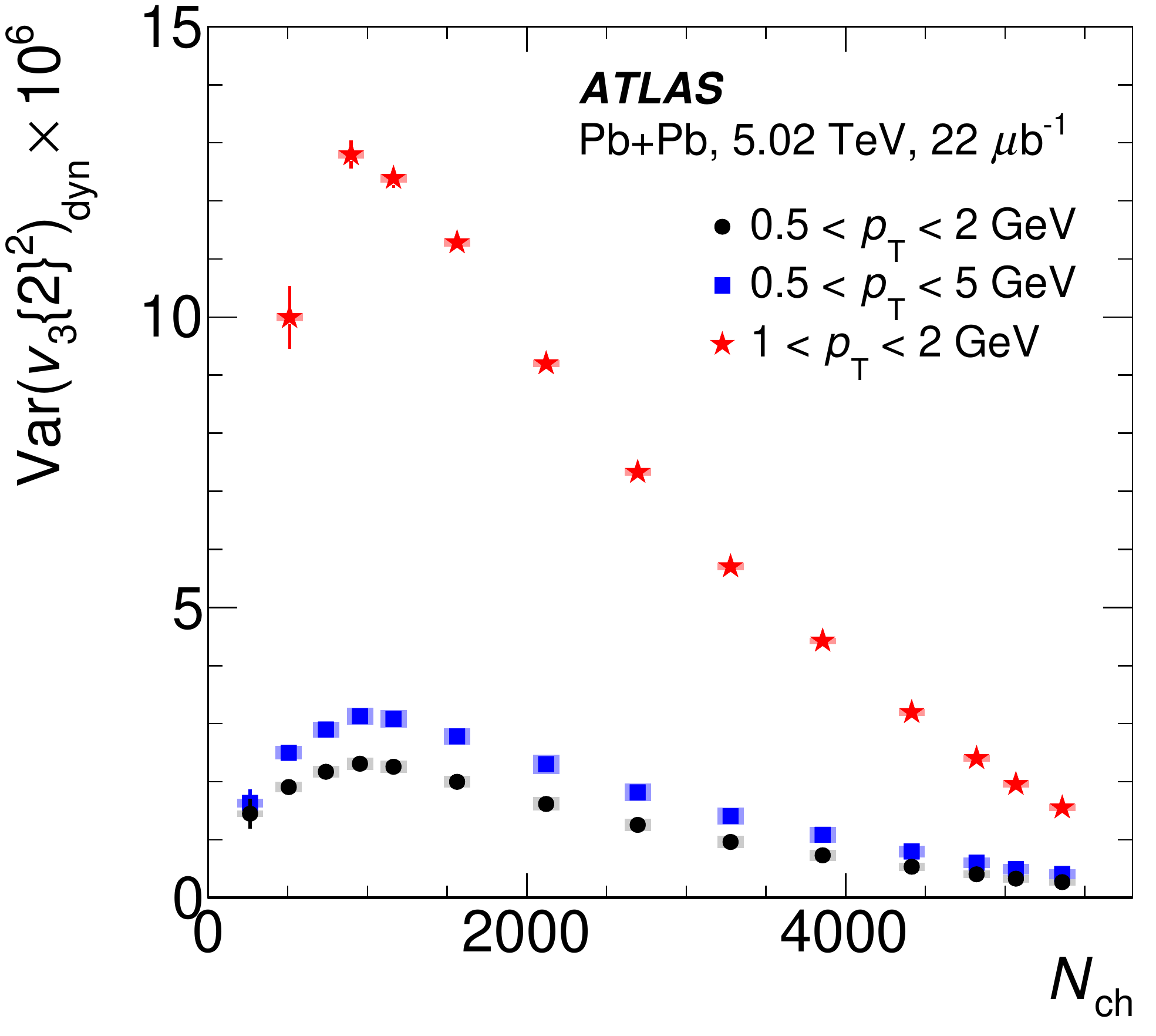}\\
\hspace{1.5cm}(a) &     \hspace{1.5cm}(b) \\
\includegraphics[width=0.45\textwidth]{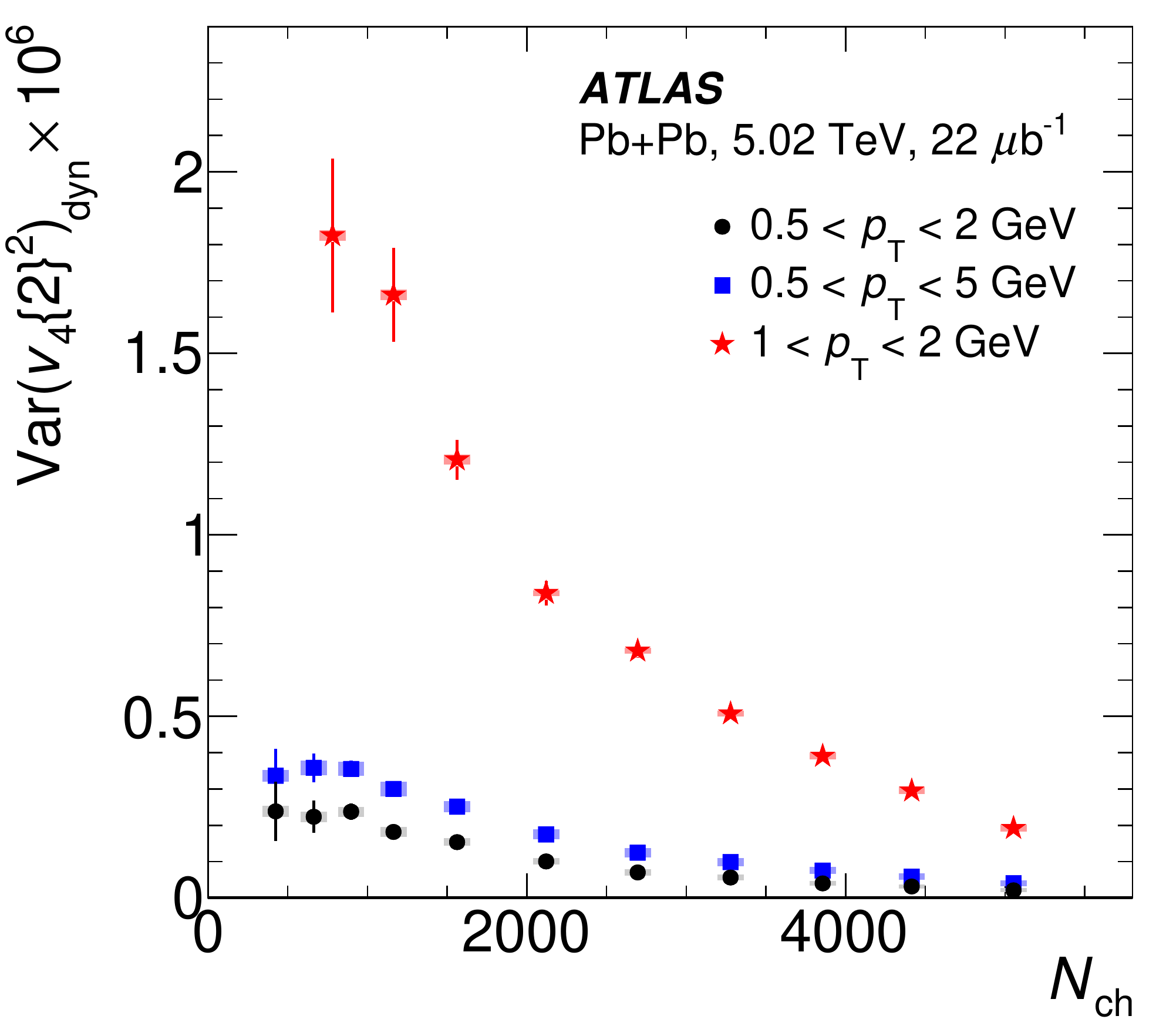}&
\includegraphics[width=0.45\textwidth]{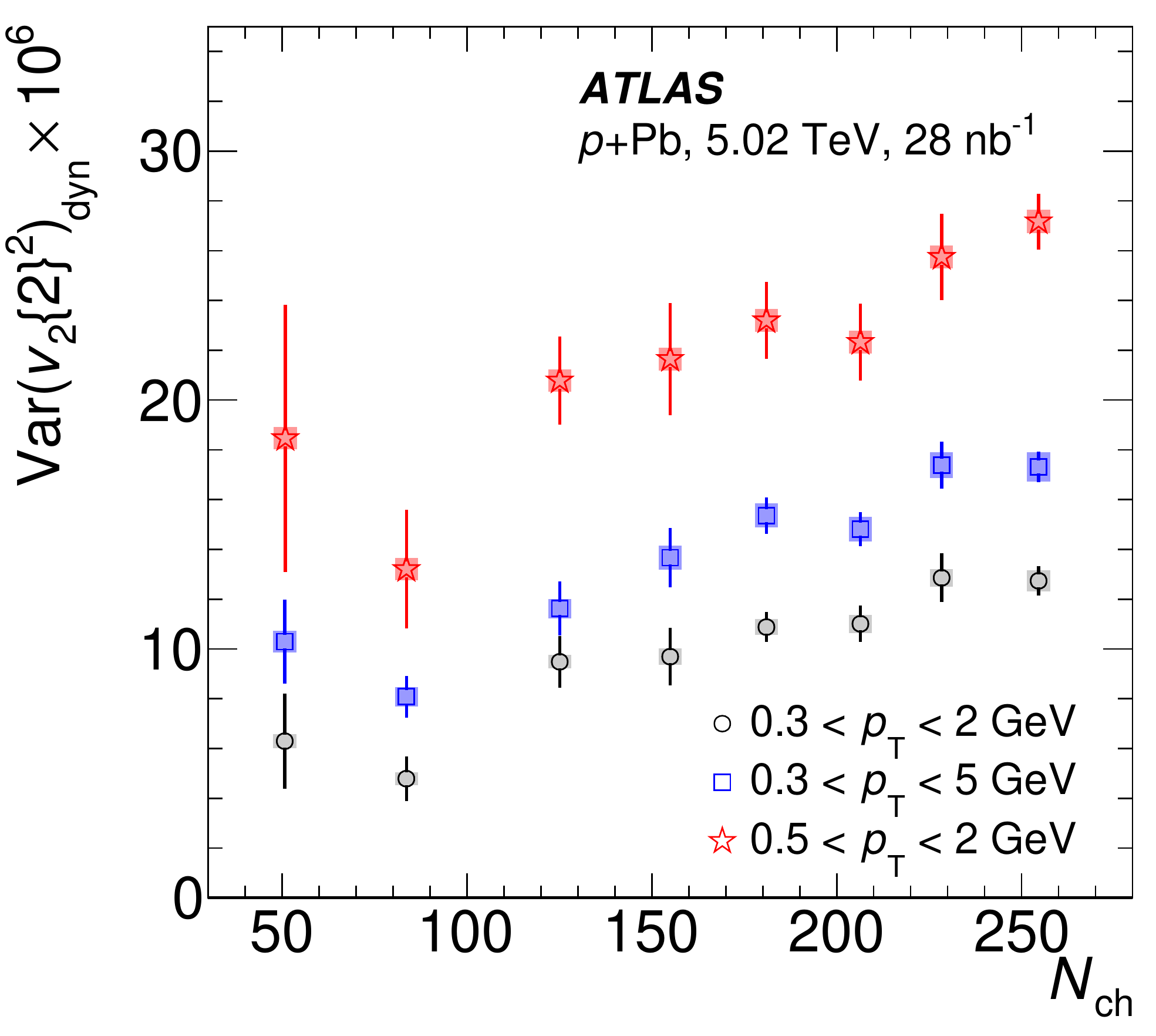}\\
\hspace{1.5cm}(c) &     \hspace{1.5cm}(d) \\
\end{tabular}
\caption{ The variance $\varn{n}$ for $n$ = 2--4 for (a)--(c) \PbPb
collisions and  $\varn{2}$ for (d) \pPb
collisions for the three \pT\ intervals as a function
of charged-particle multiplicity \Nch. The statistical and systematic
uncertainties are shown as vertical error bars and boxes, respectively.  }
\label{fig:var}
\end{figure}

Figure~\ref{fig:var} shows $\varn{n}$ for $n$ = 2--4 as function of \Nch\ for \PbPb collisions. 
For low multiplicities, $\varn{n}$ increases with increasing  $\Nch$, reaching a maximum at \Nch\ of approximately 500 (1000) for $n$ = 2 ($n$ = 3), respectively.
At higher \Nch\ values the variances decrease with multiplicity.
The dynamical variance for $n=4$, measured for $\Nch \gtrapprox 500$, decreases with increasing \Nch.
The ordering $\varn{2} > \varn{3} > \varn{4}$ and the multiplicity
dependence of  $\varn{n}$ are similar to the ordering and centrality dependence of $\vtwo{n}$ measured by ATLAS~\cite{HION-2016-06}.
Also shown in Figure~\ref{fig:var} is $\varn{2}$ for \pPb collisions as a function of  \Nch.
The dependence is monotonic, similarly to $\vtwo{2}$~\cite{HION-2017-02}.
In both collision systems and for all harmonics, the same ordering
of $\varn{n}$ depending on the \pT interval is observed.
The largest variances are observed for the \pT\ intervals with an
increased lower limit. This is expected as the $\vtwo{n}$ value increases strongly
with \pT\ below 3~\GeV~\cite{HION-2016-06}.
Additionally, the interval in which the upper limit on \pT is set to 5
\GeV\ integrates the region with the highest values of $\vtwo{n}$ (which occur around 3~\GeV) and thus the
values of the variance are expected to be larger than that for the main \pt range.
 
\begin{figure}[!htb]
\centering
\begin{tabular}{cc}
 
\includegraphics[width=0.45\textwidth]{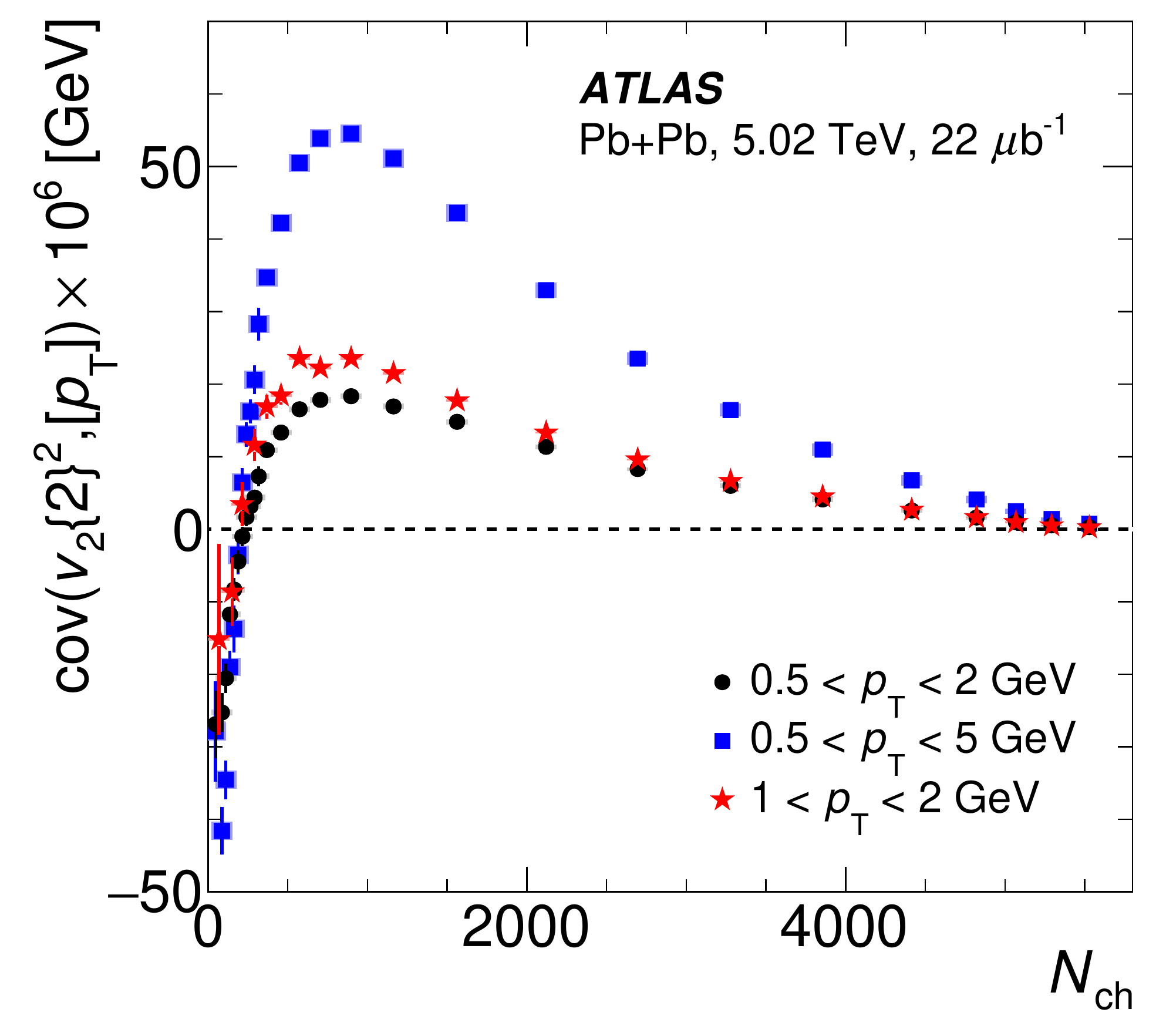}&
\includegraphics[width=0.45\textwidth]{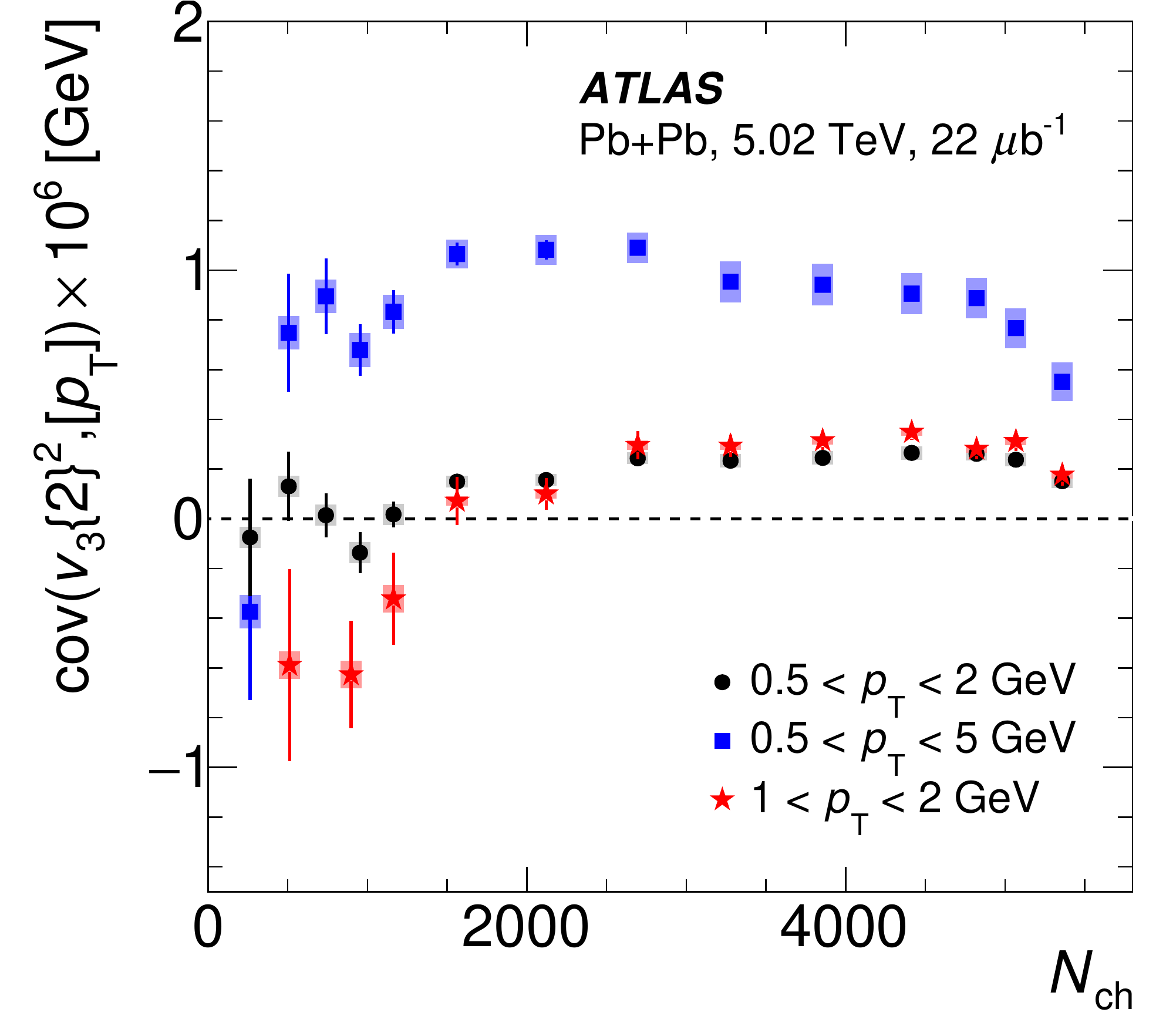}\\
\hspace{1.5cm}(a) &     \hspace{1.5cm}(b) \\
\includegraphics[width=0.45\textwidth]{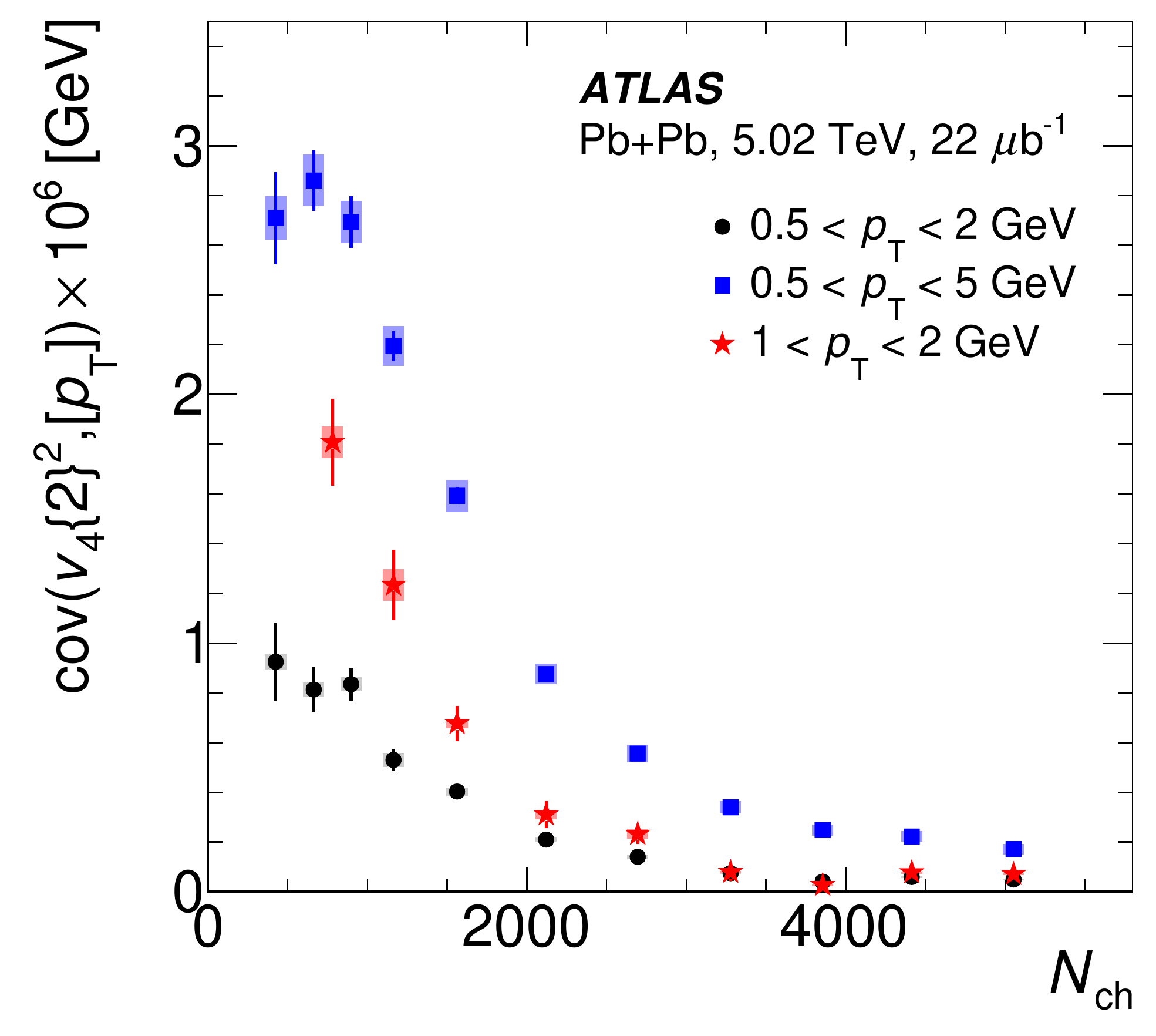}&
\includegraphics[width=0.45\textwidth]{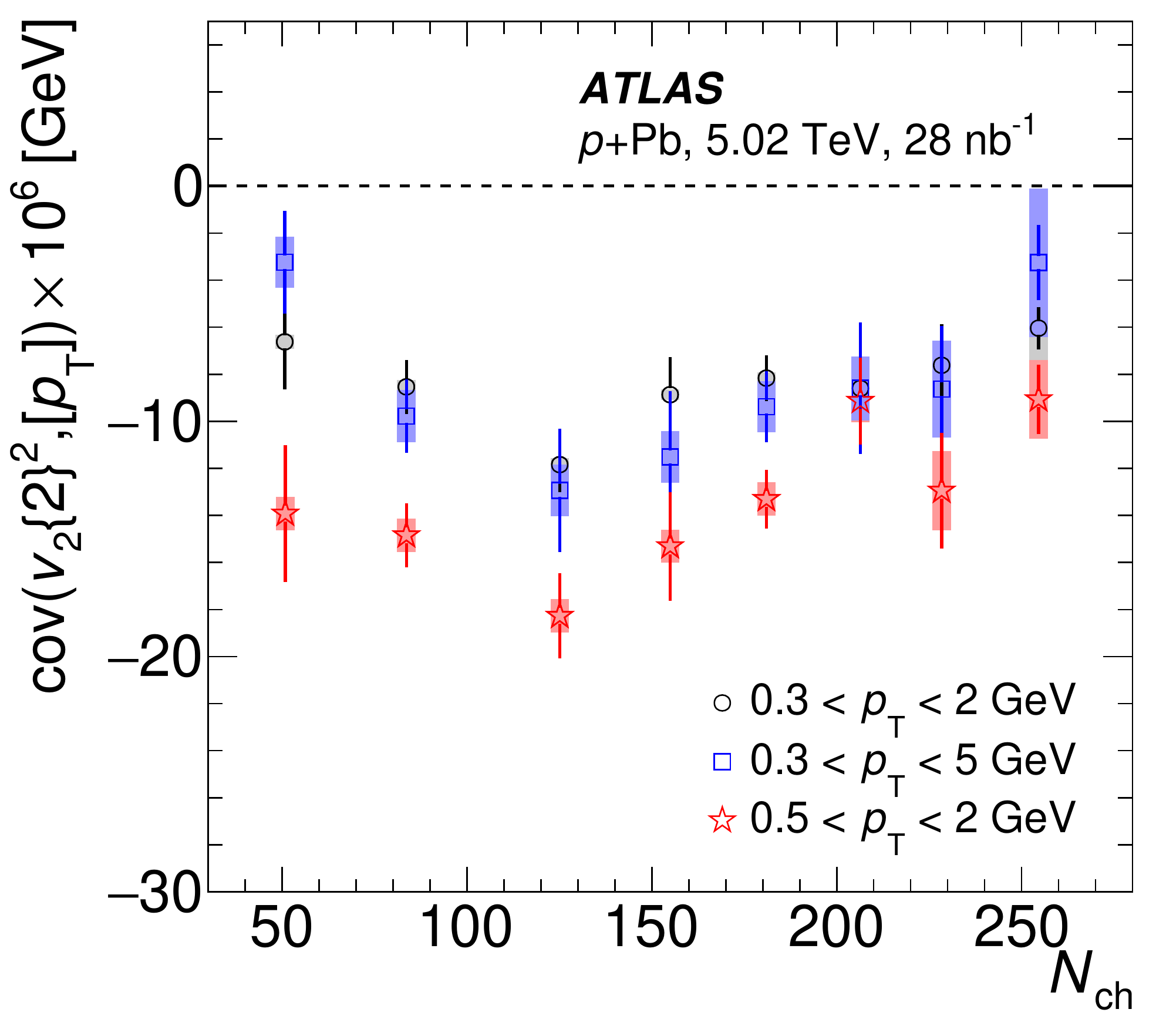}\\
\hspace{1.5cm}(c) &     \hspace{1.5cm}(d) \\
\end{tabular}
\caption{ The covariance $\covn{n}$ for $n$ = 2--4 in (a)--(c) \PbPb collisions
and $\covn{2}$ in (d) \pPb collisions for three \pT\
ranges as a function of the charged-particle multiplicity \Nch.
The statistical and systematic uncertainties are shown as vertical error bars and boxes, respectively. }
\label{fig:cov}
\end{figure}

In Figure~\ref{fig:cov}, the covariances $\covn{n}$ are shown for the 2$^{\mathrm{nd}}$-, 3$^{\mathrm{rd}}$-, and 4$^{\mathrm{th}}$-order  harmonics in \PbPb collisions and for the second-order harmonics in \pPb collisions.
They are presented as a function of \Nch\ for three \pT\ intervals.
Significant positive correlations between $\vtwo{n}$ and \mpT
are observed in the \PbPb events.
The measured covariances depend on the charged-particle multiplicity and the \pT range of the charged particles.
In \PbPb collisions, a strong dependence on the multiplicity is observed for $n=2$ and 4.
The $\covn{3}$ depends only weakly on \Nch.
A negative $\covn{2}$ is measured at multiplicities $\Nch <$ 200 and a
negative $\covn{3}$ for 1 $<\pT\ <$ 2~\GeV\ below $\Nch <$ 1800.
The covariances \covn{2}\ in \pPb\ events are negative in the entire measured \Nch\ range
and show weak \Nch\ dependence.
Unlike in \PbPb events, the \covn{2} in \pPb\ events have similar magnitudes for different \pT\ intervals.
 
\subsection{The modified \pcc}
\begin{figure}[!htb]
\centering
\begin{tabular}{cc}
\includegraphics[width=0.45\textwidth]{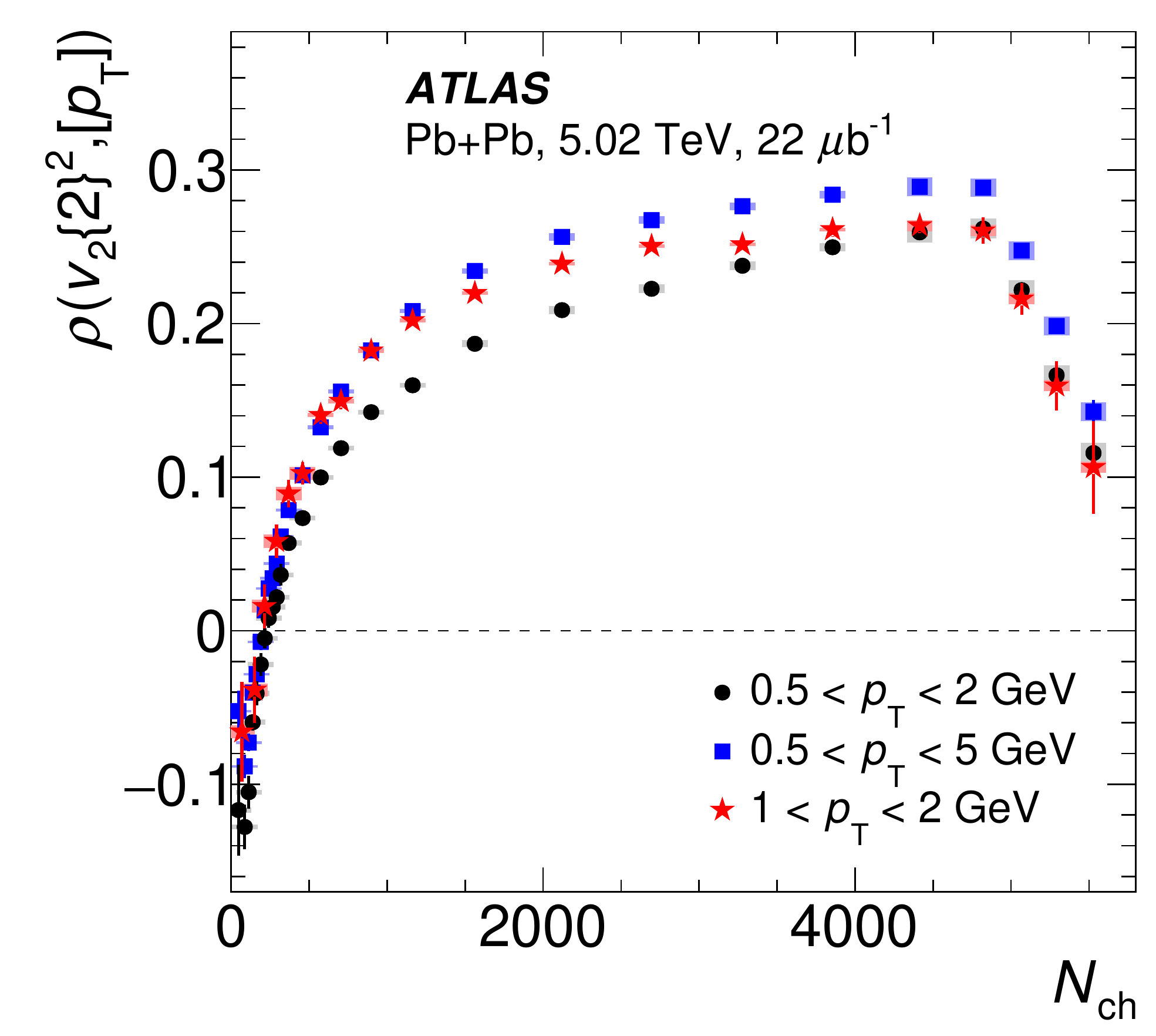}&
\includegraphics[width=0.45\textwidth]{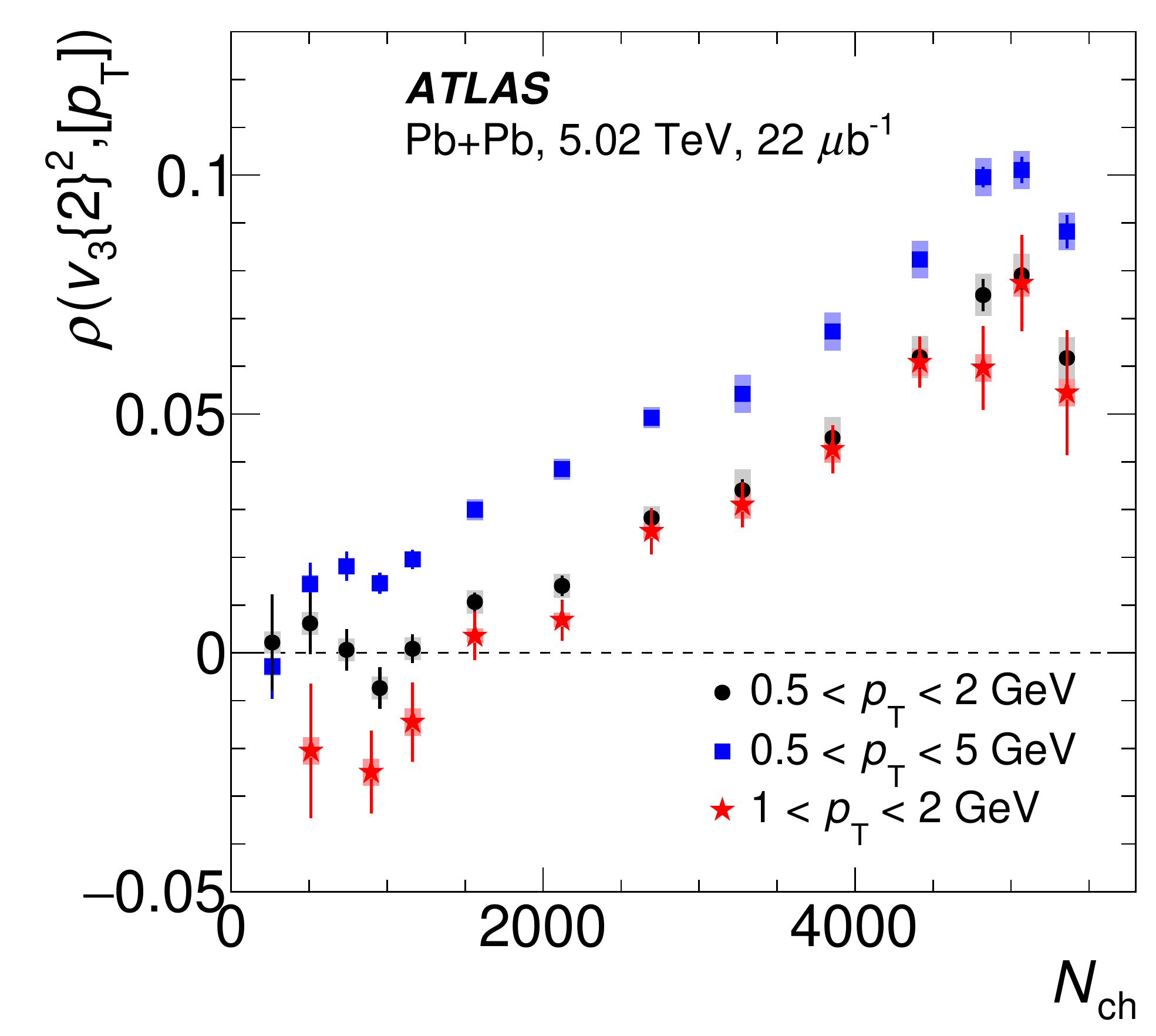}\\
\hspace{1.5cm}(a) &     \hspace{1.5cm}(b) \\
\includegraphics[width=0.45\textwidth]{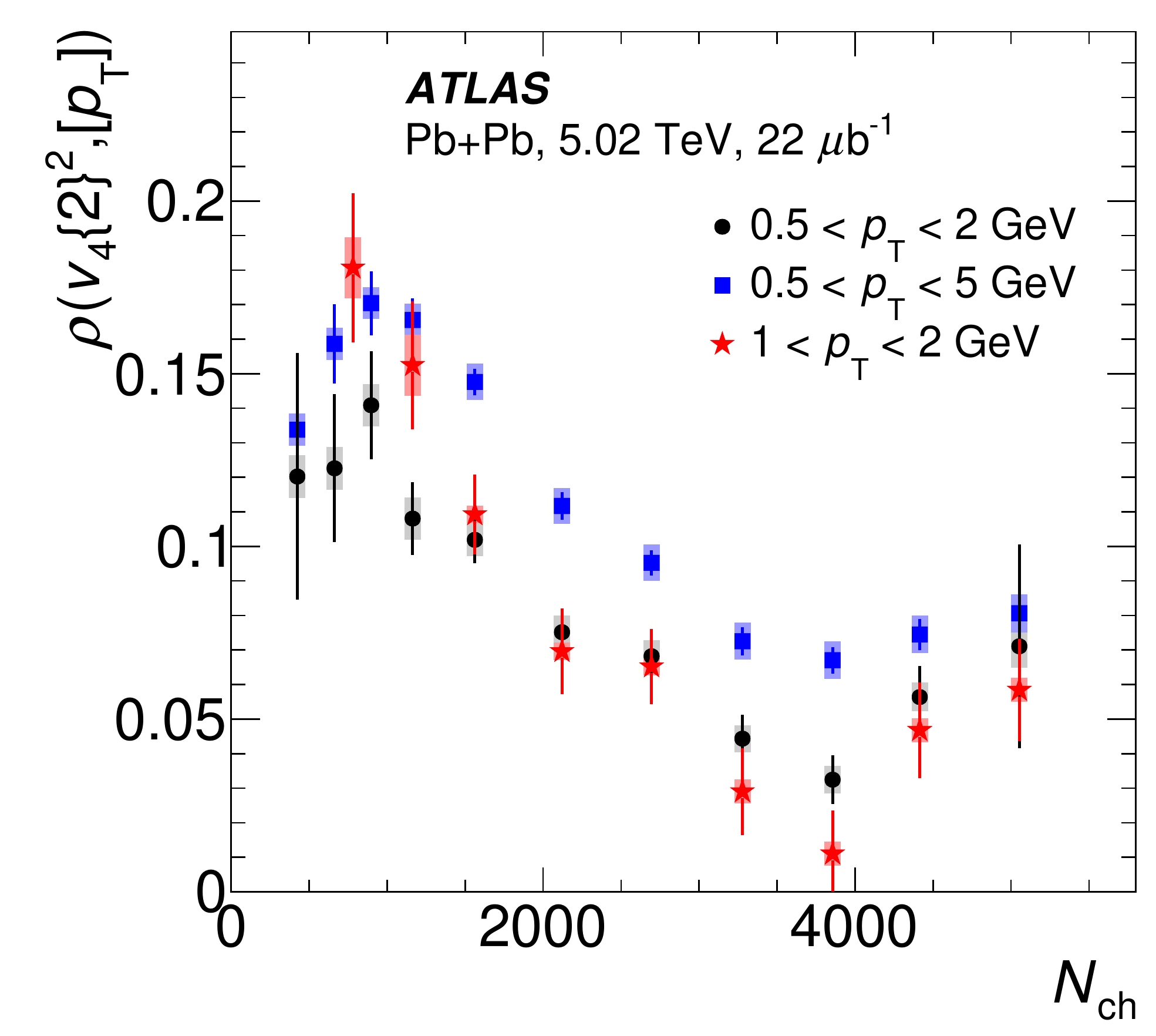}&
\includegraphics[width=0.45\textwidth]{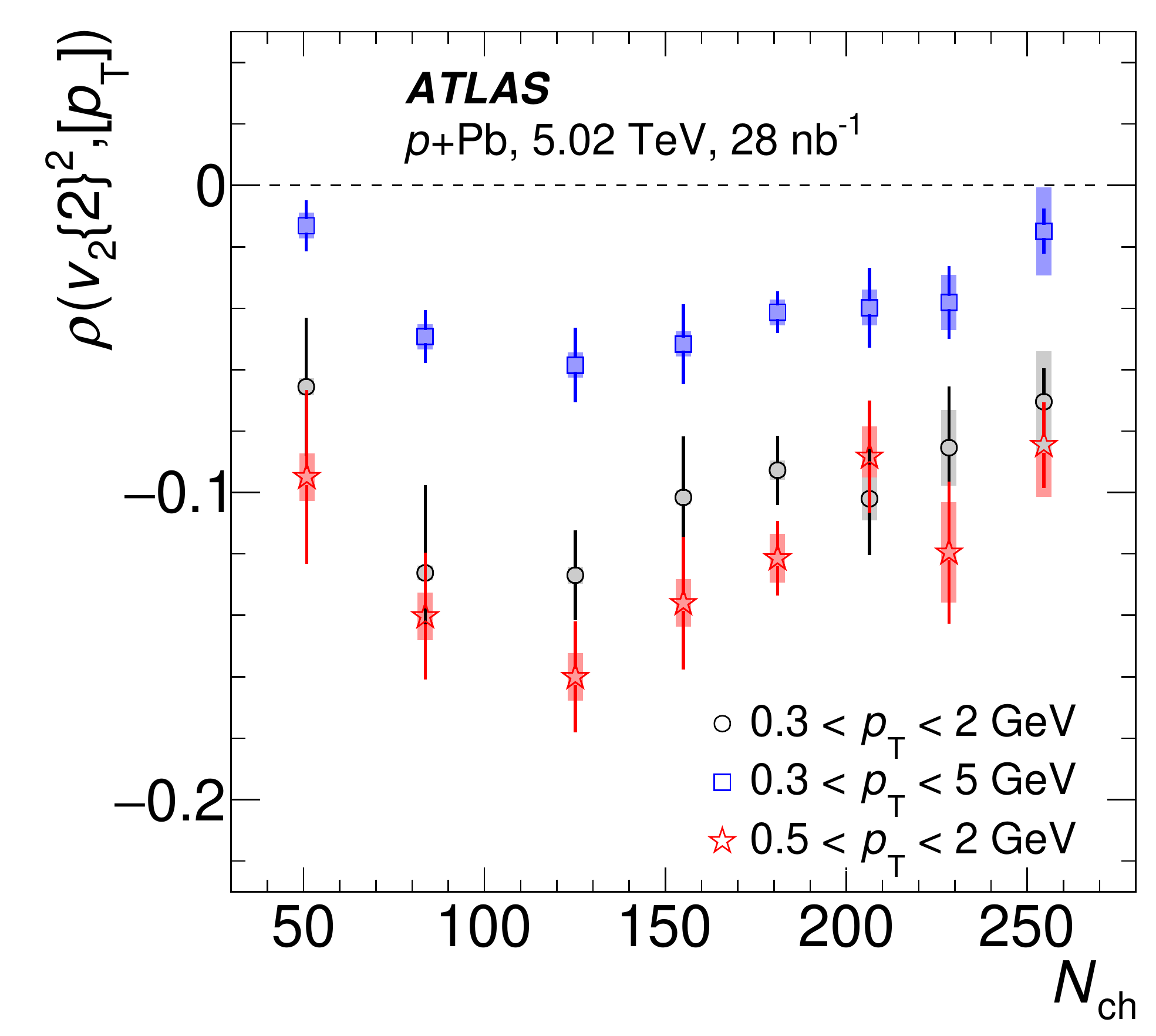}\\
\hspace{1.5cm}(c) &     \hspace{1.5cm}(d) \\
\end{tabular}
\caption{ The \pcc\ $\rhon{n})$ for $n$ = 2--4 in (a)--(c) \PbPb collisions  and
(d) \pPb collisions as a function of the charged-particle multiplicity \Nch\ for three \pT\ ranges.
The statistical and systematic uncertainties are shown as vertical error bars and boxes, respectively. }
\label{fig:rho_nch}
\end{figure}

The modified \pcc\
$\rhon{n}$ for $n$ = 2--4 in \PbPb\ collisions and for $n=2$ in \pPb\ collisions is shown in Figure~\ref{fig:rho_nch}.
In \PbPb collisions, the behaviour of \rhon{2} is similar for all \pT\ intervals.
It starts at negative values for $\Nch < 200$ and rapidly increases
with multiplicity up to $\sim$1500 particles where the increase
slows down and reaches the maximum at $\Nch\approx4500$ of 0.24--0.3, depending on the
\pT interval. At even higher \Nch, the \rhon{2} value decreases
rapidly.
The significant correlation observed for mid-central events suggests a
connection between anisotropic and radial~\cite{VPS} flows which might
be attributed to stronger hydrodynamic response (larger pressure gradients) to the large initial-state eccentricities~\cite{Mazeliauskas:2015efa}.
The modified \pcc multiplicity dependence could reflect a balance between stronger radial flow observed in central collision and the larger initial eccentricity seen in peripheral interactions.
The decrease observed in central collisions, for $\Nch \gtrsim 5000$, might be related to the increased role of initial-state fluctuations in anisotropic flow~\cite{HION-2012-04}.
However, a complete understanding of this effect would require a more precise modelling of heavy ion collisions.
The correlation coefficients calculated with the upper \pT\ limit of 2~\GeV\ are 10--20\% smaller than the values obtained with a \pT\ limit of 5~\GeV.
The correlation coefficient $\rhon{3}$ is evaluated in \PbPb\
collisions for the same three \pT\ ranges.
The magnitudes measured for \rhon{3}\ are significantly
smaller than those measured for \rhon{2}\ and similar to the
magnitudes of \rhon{4}.
All three curves increase with \Nch\ in the range of $ 1000 < \Nch < 5000 $.
At low values of $\Nch$, a flattening of the trend can be noticed.
In the most central collisions, a breakdown of the rise is seen, similarly to the $\rhon{2}$.
Above $\Nch \sim 1500$, the curves for the two intervals with the same
maximum \pT\ are consistent with each other and are below the
curve for the interval which uses tracks with \pT\ up to 5~\GeV.
The largest values of $\rhon{4}$ are observed at $\Nch\approx 1000$.
For high \Nch,
$\rhon{4}$ decreases with \Nch\ up to about
$\Nch \approx 4000$ and rises slowly at higher values.
The trends obtained for \pT\ intervals with the same minimum value are
consistent above $\Nch \sim 1500$ as is the case for $\rhon{3}$.
The decrease for $\Nch < 4000$ might be due to a contribution to $v_{\mathrm 4}$ from a non-linear term containing $v_{\mathrm 2}^2$,
decreasing with increasing centrality~\cite{HION-2014-03}.
However, a theoretical modelling of the initial state and its subsequent evolution would be required to support this interpretation.
Similarly to the \rhon{3}, the \rhon{4} correlations measured with the larger
upper \pt\ limit have larger magnitudes.
The results for the larger upper \pt\ limit show the sensitivity of
the $\rhon{n}$ coefficients to the high \pT\ part of the particle spectrum contaminated with non-flow correlations from jets.
On the other hand, the correlations measured for the intervals with
fixed upper \pT limit (2 \GeV) and  varied lower \pT limits are similar,
demonstrating insensitivity of the modified \pcc\ coefficients to a significant change of the event charged-particle multiplicity as expected~\cite{Bozek:2016yoj}.
The fourth-order correlations are weaker than those for the second-order
flow harmonic and for $\Nch >4000$ are comparable to $\rhon{3}$.
The results for all harmonics indicate a change in the trend in events with high \Nch\ around 4500, which suggests a change in the nature of the correlations in those events~\cite{Mazeliauskas:2015efa}.
 
In \pPb collisions,  $\rhon{2}$ exhibits much weaker \Nch\ dependence than that in \PbPb\ collisions.
For the main \pT\ interval, the modified \pcc\ assumes a negative value of approximately $-0.1$ and is almost constant within uncertainties.
Values for different lower \pT limits are similar, and the $\rhon{2}$ magnitudes for the larger upper \pt limit are smaller.
The magnitude (and sign) of the modified \pcc\ in \pPb\ collisions is
expected to be related to the distribution of the energy
deposition in the initial state, as predicted by the hydrodynamic
model~\cite{Bozek:2016yoj}.
In hydrodynamics,  in \pPb\ collision, for small sources a higher initial pressure gradients
and smaller eccentricities are expected to be generated.
This mechanism could lead to the negative correlation of the final
state observables, this is the mean transverse momentum and higher order flow harmonics.
Thus, the negative value of the modified \pcc\ for $\vtwo{2}$ in \pPb\ and peripheral \PbPb\ that is measured should provide valuable
constraints for models describing the collectivity in small systems.
 
\subsection{Comparison of \pPb and \PbPb results}
\begin{figure}[!h]
\centering
\begin{tabular}{cc}
 
\includegraphics[width=0.45\textwidth]{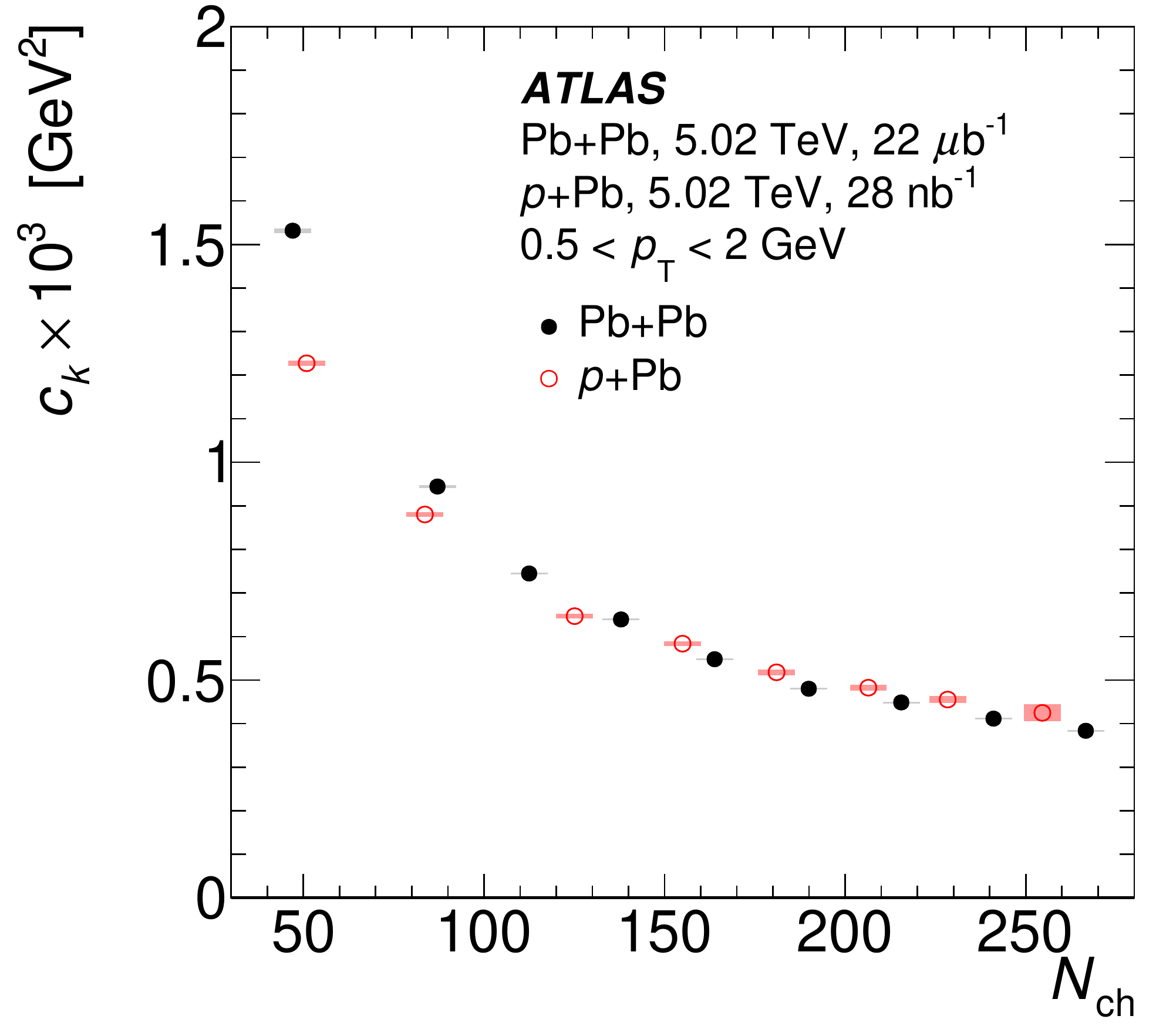} &
\includegraphics[width=0.45\textwidth]{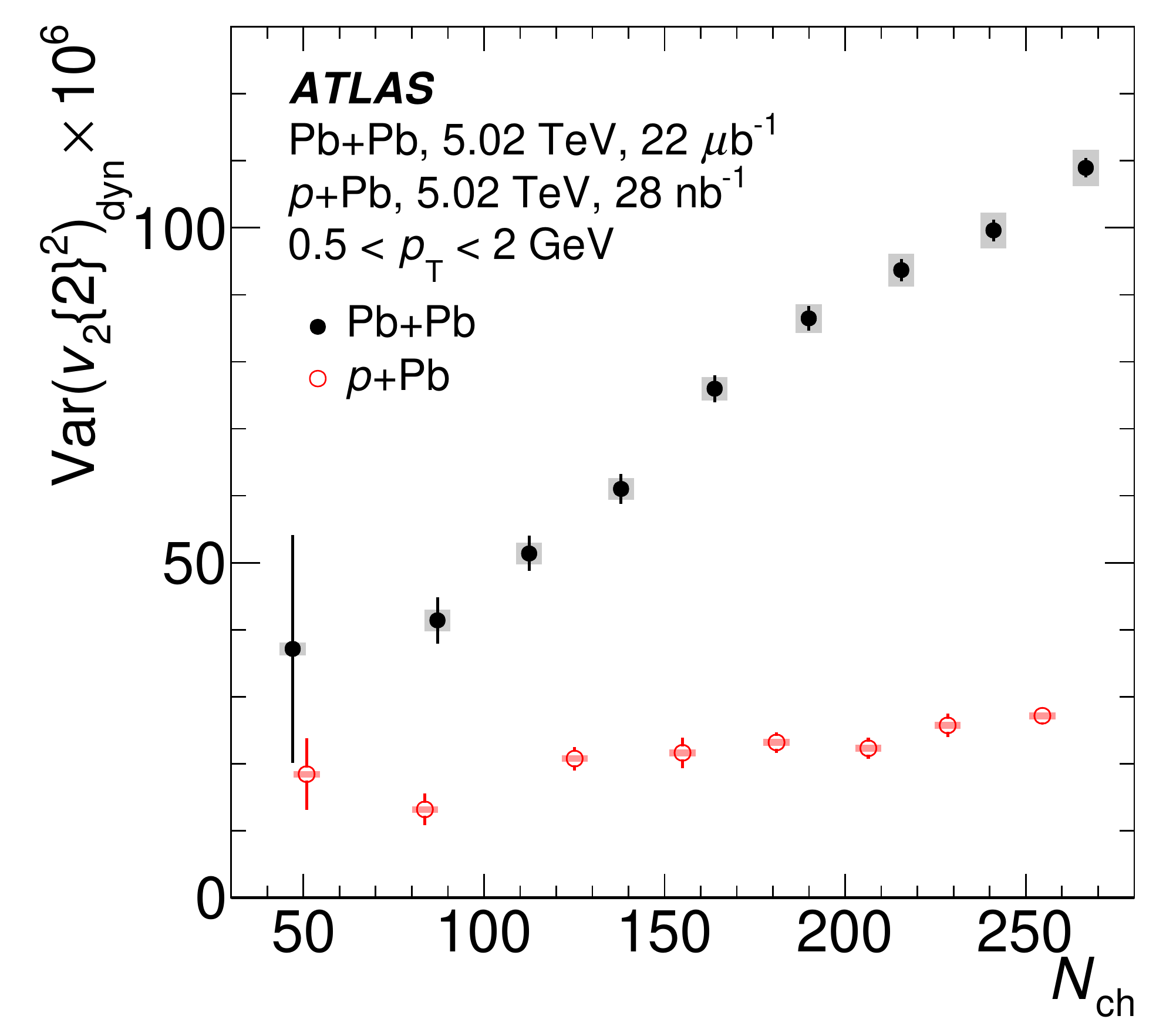}\\
\hspace{1.5cm}(a) &     \hspace{1.5cm}(b) \\
 
\includegraphics[width=0.45\textwidth]{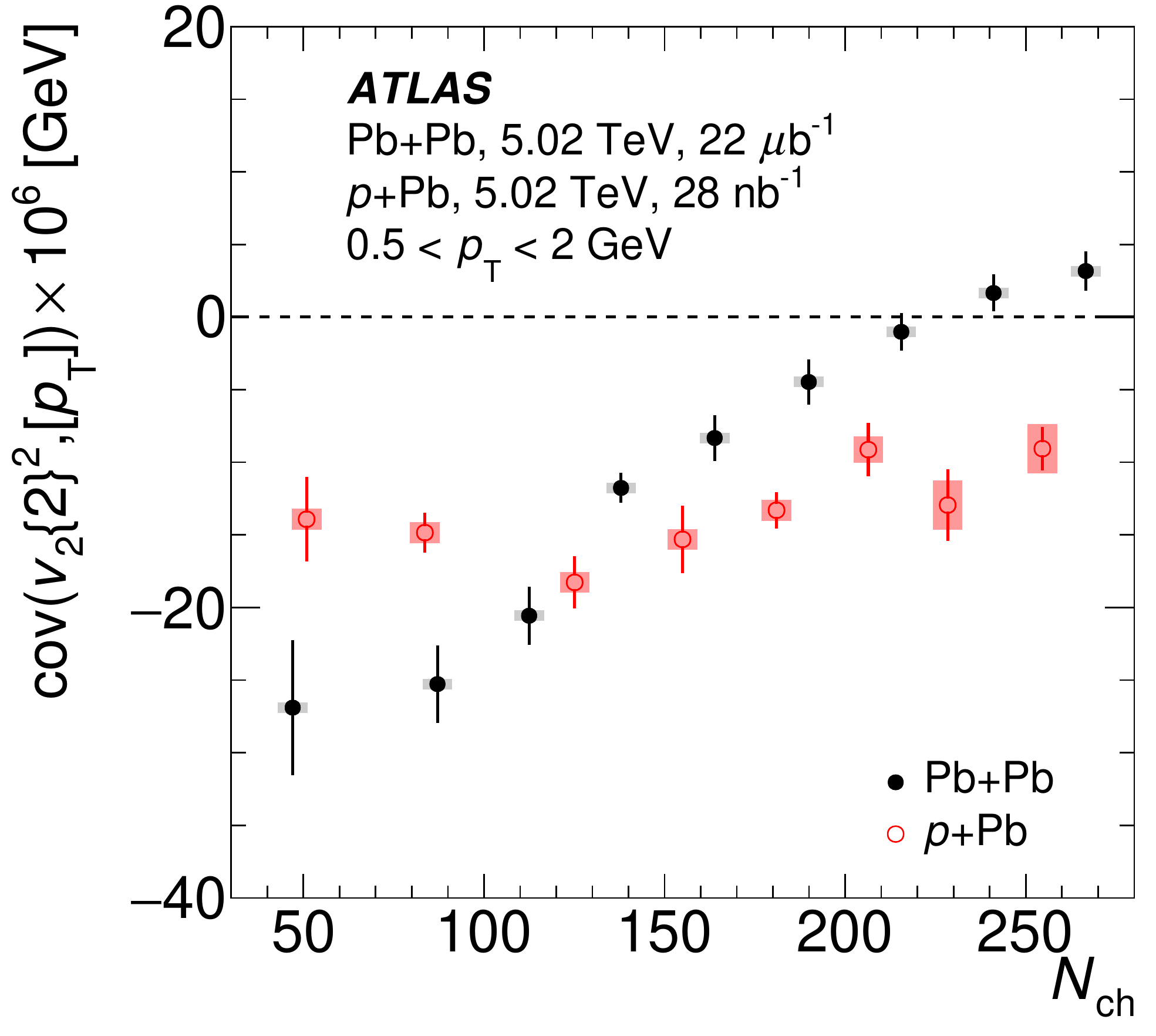}&
\includegraphics[width=0.45\textwidth]{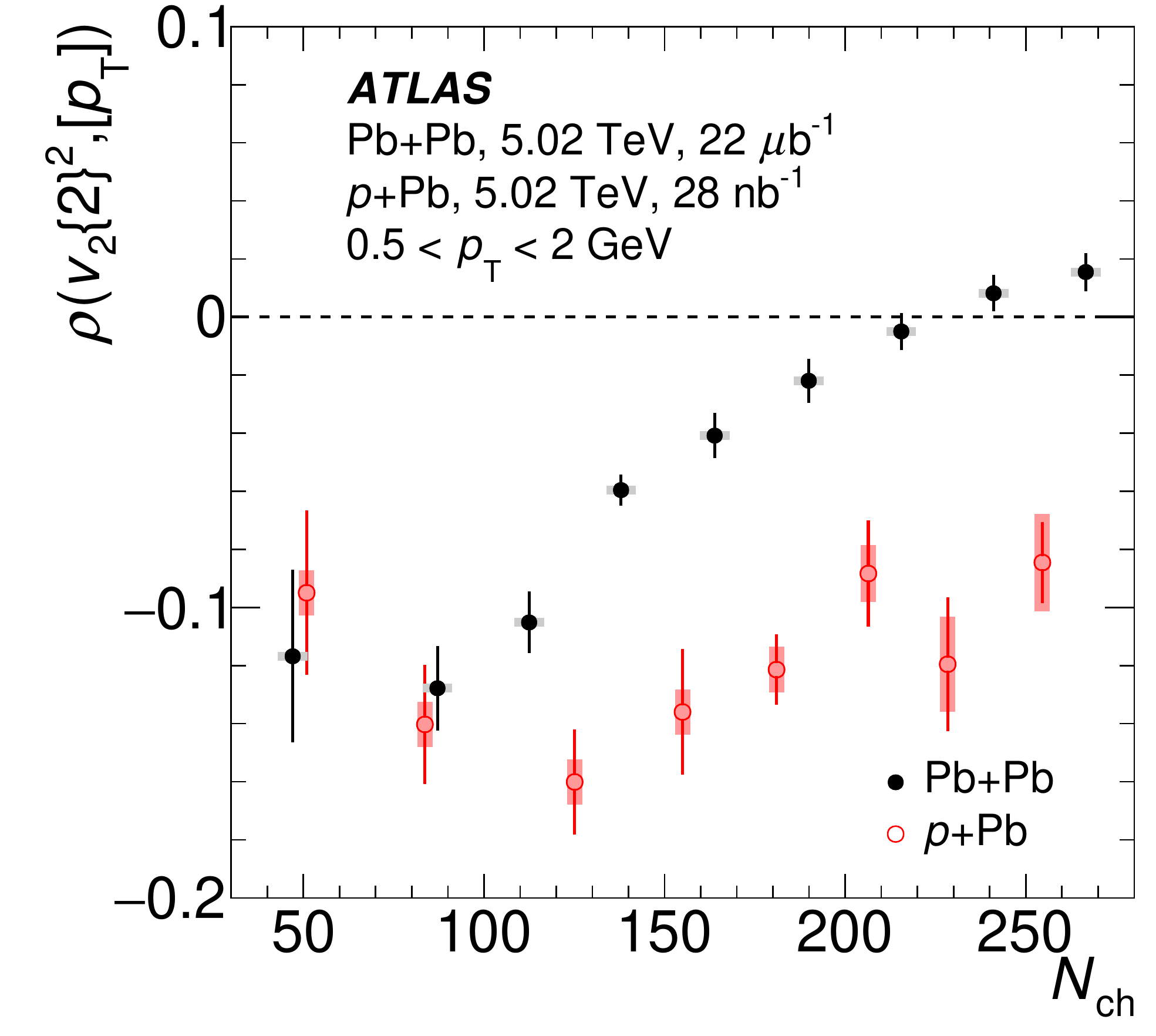}\\
\hspace{1.5cm}(c) &     \hspace{1.5cm}(d) \\
\end{tabular}
 
\caption{Comparison of (a) \ck, (b)  $\varn{2}$,
(c) $\covn{2}$, and the (d) $\rhon{2}$
for the range $0.5 < \pT < 2~\GeV$ as a function
of the charged-particle multiplicity \Nch.  The statistical and systematic
uncertainties are shown as vertical error bars and boxes,
respectively.}
\label{fig:data-comp}.
\end{figure}
Figure~\ref{fig:data-comp} shows a comparison of \pPb and \PbPb
results shown in Figs.~\ref{fig:ck}$-$\ref{fig:rho_nch} for the common
\pT interval of $0.5 < \pT < 2~\GeV$.
The values of the \ck\ (Figure~~\ref{fig:data-comp}(a)) are similar
for \pPb\ and \PbPb\ collisions in this \pT\
interval, while the behaviour of the dynamical variance $\varn{2}$
(Figure~~\ref{fig:data-comp}(b)) is very different due to the
different initial eccentricities in the overlap regions in \PbPb and \pPb collisions.
Only a small rise with the multiplicity is
observed for \pPb\ collisions, which is in agreement with a slow increase of $\vtwo{2}$
with growing event
activity~\cite{HION-2015-14,HION-2017-02,HION-2013-04}.
For $\Nch\approx  50$, the dynamical variances are comparable between \PbPb\ and \pPb\ collisions.
The \Nch\ dependence of \covn{2}  is significantly different for \PbPb\
and \pPb\ collisions. A steady rise from negative to positive values
with \Nch\ is observed for peripheral \PbPb\ collisions,
and approximately constant values are obtained
for \pPb collisions.  The \Nch\ dependence of $\rhon{2}$ is different for the
two collision systems.
Much weaker \Nch\ dependence of modified \pcc\ is observed in \pPb collisions compared to \PbPb\ collisions.
For $\Nch\ <100 $ the values of
$\rhon{2}$ are consistent between \PbPb\ and \pPb\ collisions.
The negative $\rhon{2}$ coefficients for the small systems in \pPb and \PbPb
collisions may suggest a more compact source model~\cite{Bozek:2016yoj}.
The comparison of the systems underlines the importance of the initial
stage in the correlations described by the $\rhon{2}$ coefficient.
The theoretical predictions for midcentral and central \PbPb\
collisions suggests that for a large system an increase of the mean
transverse momentum indicates a stronger transverse flow and a
stronger collective response to the initial geometry of the source,
characterized by the positive value of the modified \pcc.
 
\subsection{Comparison to theoretical predictions}
\begin{figure}[!htb]
\begin{center}
\begin{tabular}{cc}
\includegraphics[width=0.45\textwidth]{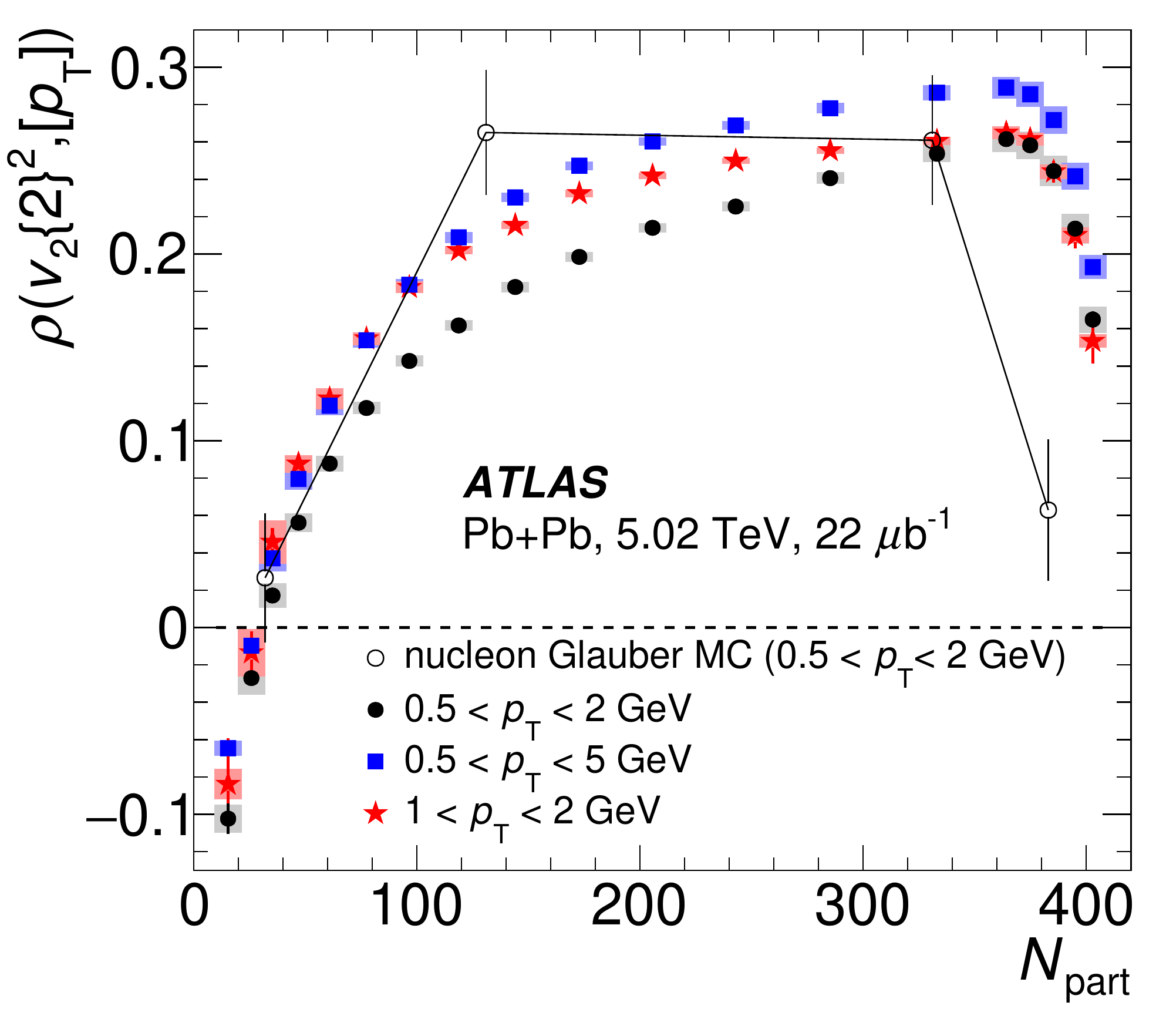}&
\includegraphics[width=0.45\textwidth]{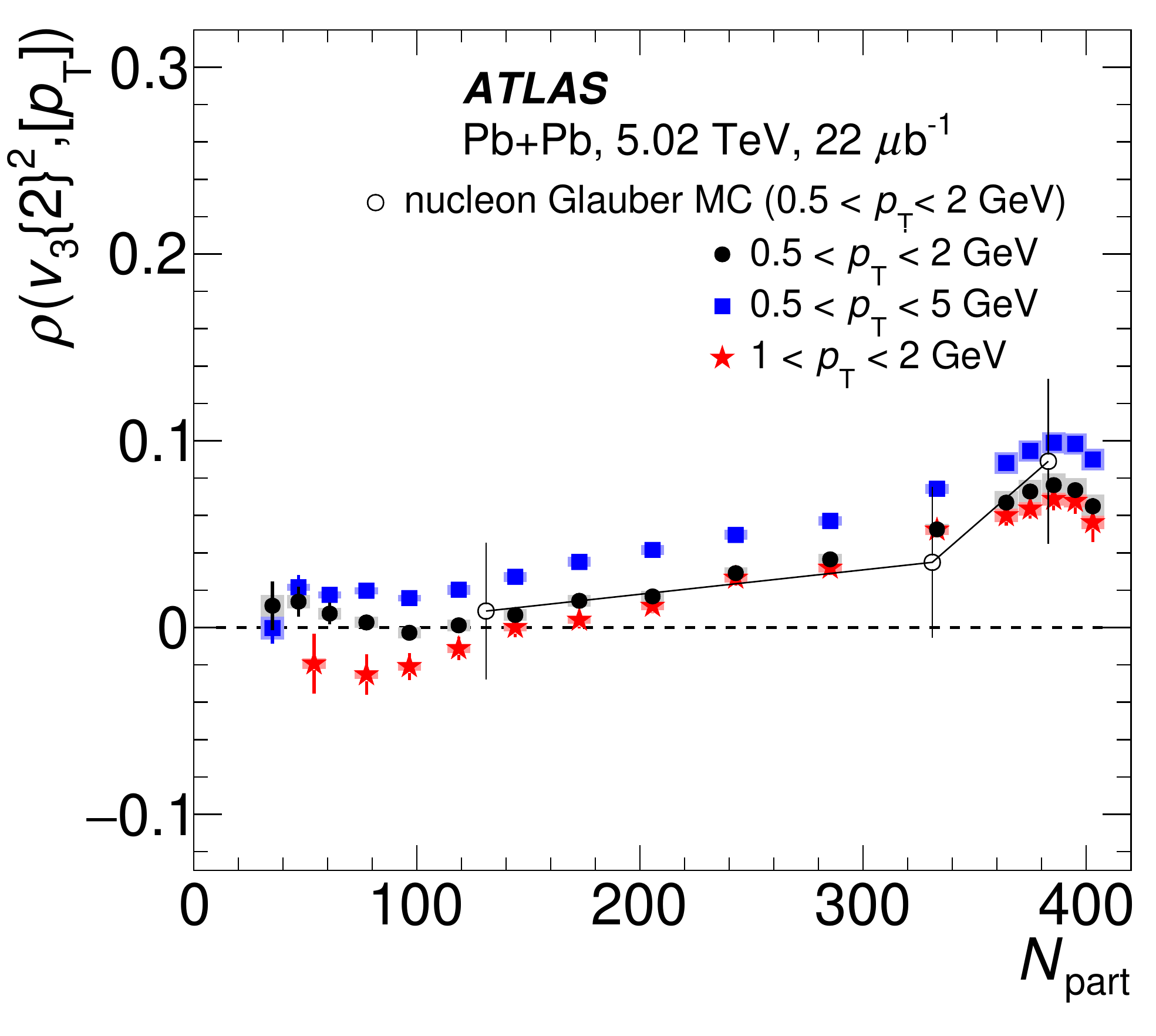}\\
\hspace{1.5cm}(a) &     \hspace{1.5cm}(b) \\
\end{tabular}
\begin{tabular}{c}
\includegraphics[width=0.45\textwidth]{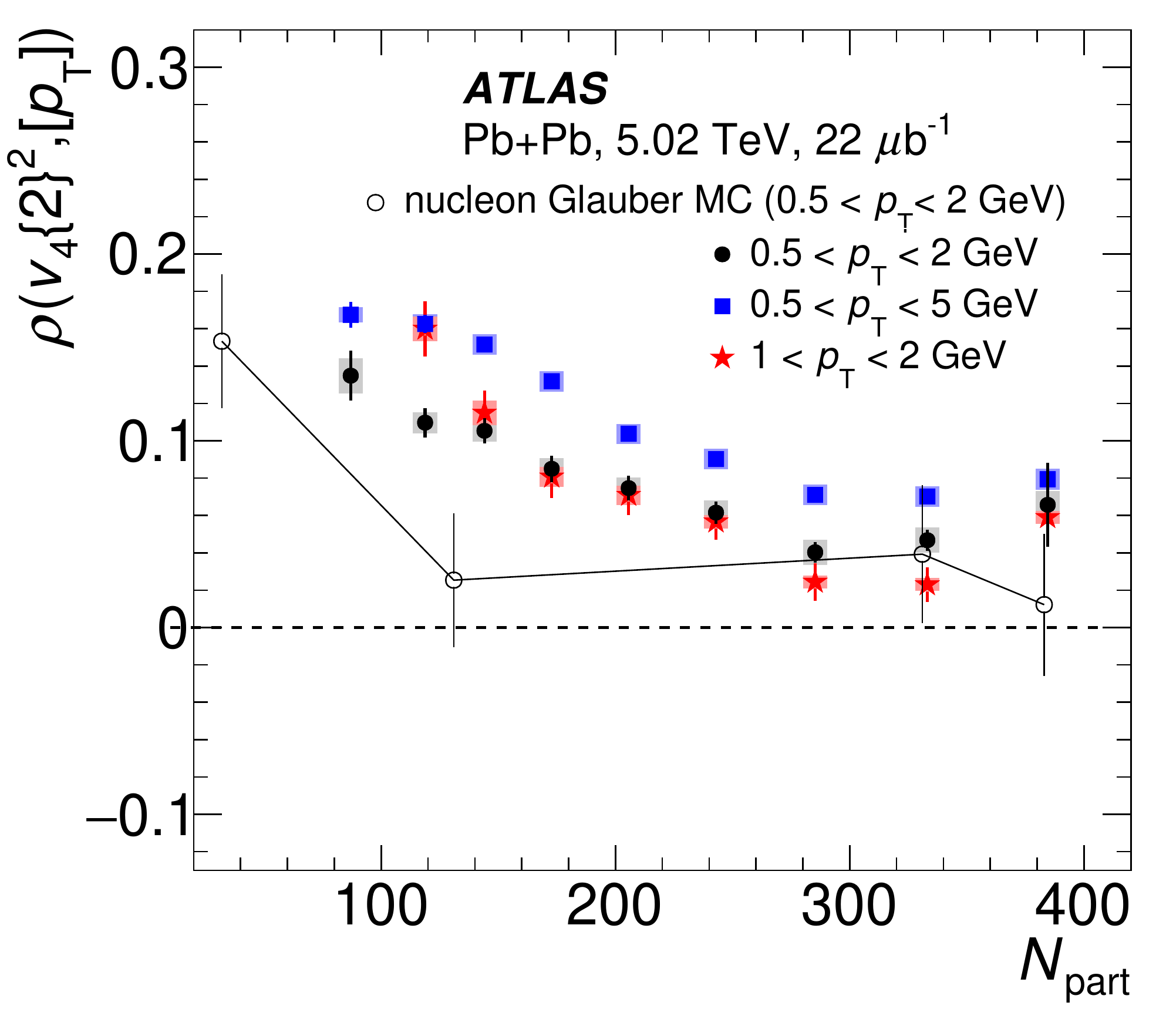} \\
\hspace{1cm}(c) \\
\end{tabular}
\end{center}
\vspace{-0.5cm}
\caption{ The \pcc\ $\rho( \vnpT{n} )$ for (a) $n$ = 2, (b) $n$ = 3, and (c) $n$ = 4 in \PbPb
collisions as a function of $\Npart$ for three \pt ranges. The statistical and systematic
uncertainties are shown as vertical error bars and boxes,
respectively. A comparison with model
predictions~\cite{Bozek:2016kpf} is also shown with a line added
to guide the eye.}
\label{fig:rho_npart}
\end{figure}

To compare the \PbPb\ results with a theoretical prediction in Ref.~\cite{Bozek:2016yoj},
the $\rhon{n}$ coefficients for $0.5 < \pT < 2$~\GeV\ are obtained as a function
of centrality intervals expressed by $\Npart$ using the procedure described in
Section~\ref{sec:analysis}.
Figure~\ref{fig:rho_npart} shows the
$\Npart$ dependence of $\rhon{n}$ for $n$ = 2--4 in \PbPb collisions.
It resembles the trends observed in Fig.~\ref{fig:rho_nch}, which show the modified \pcc\ as a function of $\Nch$, a measure of event activity.
The theoretical predictions of the $\rhon{n}$ coefficient are based on
a  model in which the initial conditions were generated with nucleon positions by a MC Glauber model~\cite{Loizides:2016djv}.
These initial conditions are then evolved using the pressure-driven 3+1D hydrodynamical
simulations with viscous effects followed by the statistical
particle emission to match multiplicities observed
experimentally~\cite{Bozek:2016kpf}.
The modified Pearson correlation coefficient is then extracted from
the final-state particles.
The predictions for all harmonics are  consistent with the data within
the large model uncertainties
except for the most central collisions where the predictions
underestimate the measured $\rhon{2}$ and for the semi-peripheral
collisions, for $\Npart \sim 130$, where the predictions overestimate
the $\rhon{2}$ and underestimate $\rhon{4}$.
 
\FloatBarrier
 
\FloatBarrier

\section{Summary}
\label{sec:summary}
 
The first measurement of the modified \pcc\
$\rhon{n}$, which quantifies the correlation between the flow harmonics and the mean transverse
momentum, is performed by ATLAS experiment at the LHC.
The measurement uses  $22~\imub$ of
\PbPb data and $28~\inb$ of \pPb data at the same centre-of-mass
energy per nucleon pair of 5.02~\TeV.
 
The correlation coefficient for several charged-particle \pT\ ranges is measured as a function of the number of
charged particles \Nch  and, in \PbPb\ collisions, the average number of nucleons participating in the collision,
\Npart.
For the 2$^{\mathrm{nd}}$-, 3$^{\mathrm{rd}}$-, and 4$^{\mathrm{th}}$-order harmonics, the measured
quantities exhibit a dependence on the choice of charged-particle \pT\ range.
Measurements with an upper limit of 5~\GeV\ on \pt\  indicate a stronger correlation
than those with an upper limit of 2~\GeV.
For mid-central and central collisions, when varying the lower \pT\
limit,
consistent  values of  $\rhon{3}$ and $\rhon{4}$ coefficients are
obtained,
whereas for the $\rhon{2}$ coefficient a difference of 10--20\% is
seen.
As a function of event activity, for \PbPb\ collisions, a strong
positive correlation $\rhon{2}$
is observed in mid-central and central collisions while negative
values are measured for  peripheral events.
The correlation $\rhon{3}$ is found
to be weaker, yet non-zero.
The values of $\rhon{4}$ are also positive in the
studied centrality range.
Non-monotonic behaviour is observed in central \PbPb\ collisions.
That trend observed for $\rhon{2}$ in \PbPb\ collisions is in line
with expectations drawn from the ALICE results~\cite{PhysRevC.93.034916}.
In \pPb collisions, the value of $\rhon{2}$ is negative and
approximately independent of \Nch.
 
The modified \pcc\ is a valuable tool for studying the dynamics of heavy-ion collisions.
It provides a reliable estimate of the magnitude of correlations
calculated using finite multiplicities.
In comparison with existing results, it allows quantitative comparisons between the experimental data and theoretical models.
The precise measurements of this observable, presented in this paper, provide useful insights into the interplay of the
azimuthal anisotropies (azimuthal flow) and the mean event \pT (radial
flow), providing input for a better understanding of QGP dynamics and for constraining the theoretical models.
The obtained $\rhon{n}$ coefficients for $0.5 < \pT < 2$~\GeV\ were compared with a theoretical prediction based on the pressure-driven 3+1D hydrodynamical simulations with viscous effects. The predictions for all harmonics are consistent with the data within
the large model uncertainties. The only exception are the most central collisions, where the predictions
underestimate the measured $\rhon{2}$ and the semi-peripheral collisions, where the predictions overestimate
the $\rhon{2}$ and underestimate $\rhon{4}$.
Sizeable positive correlations observed for non-peripheral \PbPb\
collisions support a qualitatively expected scenario
in which the azimuthal flow originates from the pressure gradients.
 
In small system collisions the magnitude of the transverse flow is
expected to be very sensitive to the size of the initial source in the
hydrodynamic model.
In particular, in the compact source scenario in \pPb\ collisions, the smaller source sizes are expected to yield larger  transverse flow and smaller initial eccentricities.
The negative sign of the modified \pcc\ measured in \pPb\ collisions seems to support the compact source scenario, and indicates the role  of the initial conditions in these systems.
 
\section*{Acknowledgements}
 

We thank CERN for the very successful operation of the LHC, as well as the
support staff from our institutions without whom ATLAS could not be
operated efficiently.
 
We acknowledge the support of ANPCyT, Argentina; YerPhI, Armenia; ARC, Australia; BMWFW and FWF, Austria; ANAS, Azerbaijan; SSTC, Belarus; CNPq and FAPESP, Brazil; NSERC, NRC and CFI, Canada; CERN; CONICYT, Chile; CAS, MOST and NSFC, China; COLCIENCIAS, Colombia; MSMT CR, MPO CR and VSC CR, Czech Republic; DNRF and DNSRC, Denmark; IN2P3-CNRS and CEA-DRF/IRFU, France; SRNSFG, Georgia; BMBF, HGF and MPG, Germany; GSRT, Greece; RGC and Hong Kong SAR, China; ISF and Benoziyo Center, Israel; INFN, Italy; MEXT and JSPS, Japan; CNRST, Morocco; NWO, Netherlands; RCN, Norway; MNiSW and NCN, Poland; FCT, Portugal; MNE/IFA, Romania; MES of Russia and NRC KI, Russia Federation; JINR; MESTD, Serbia; MSSR, Slovakia; ARRS and MIZ\v{S}, Slovenia; DST/NRF, South Africa; MINECO, Spain; SRC and Wallenberg Foundation, Sweden; SERI, SNSF and Cantons of Bern and Geneva, Switzerland; MOST, Taiwan; TAEK, Turkey; STFC, United Kingdom; DOE and NSF, United States of America. In addition, individual groups and members have received support from BCKDF, CANARIE, Compute Canada and CRC, Canada; ERC, ERDF, Horizon 2020, Marie Sk{\l}odowska-Curie Actions and COST, European Union; Investissements d'Avenir Labex, Investissements d'Avenir Idex and ANR, France; DFG and AvH Foundation, Germany; Herakleitos, Thales and Aristeia programmes co-financed by EU-ESF and the Greek NSRF, Greece; BSF-NSF and GIF, Israel; CERCA Programme Generalitat de Catalunya and PROMETEO Programme Generalitat Valenciana, Spain; The Royal Society and Leverhulme Trust, United Kingdom.
 
The crucial computing support from all WLCG partners is acknowledged gratefully, in particular from CERN, the ATLAS Tier-1 facilities at TRIUMF (Canada), NDGF (Denmark, Norway, Sweden), CC-IN2P3 (France), KIT/GridKA (Germany), INFN-CNAF (Italy), NL-T1 (Netherlands), PIC (Spain), ASGC (Taiwan), RAL (UK) and BNL (USA), the Tier-2 facilities worldwide and large non-WLCG resource providers. Major contributors of computing resources are listed in Ref.~\cite{ATL-GEN-PUB-2016-002}.
 

\clearpage
\appendix
\part*{Appendix}
\addcontentsline{toc}{part}{Appendix}
\section{ Systematic uncertainty of $\ck$, $\varn{n}$  and $\covn{n}$ }
\begin{figure}[!htb]
\centering
\begin{tabular}{cc}
\includegraphics[width=0.45\textwidth]{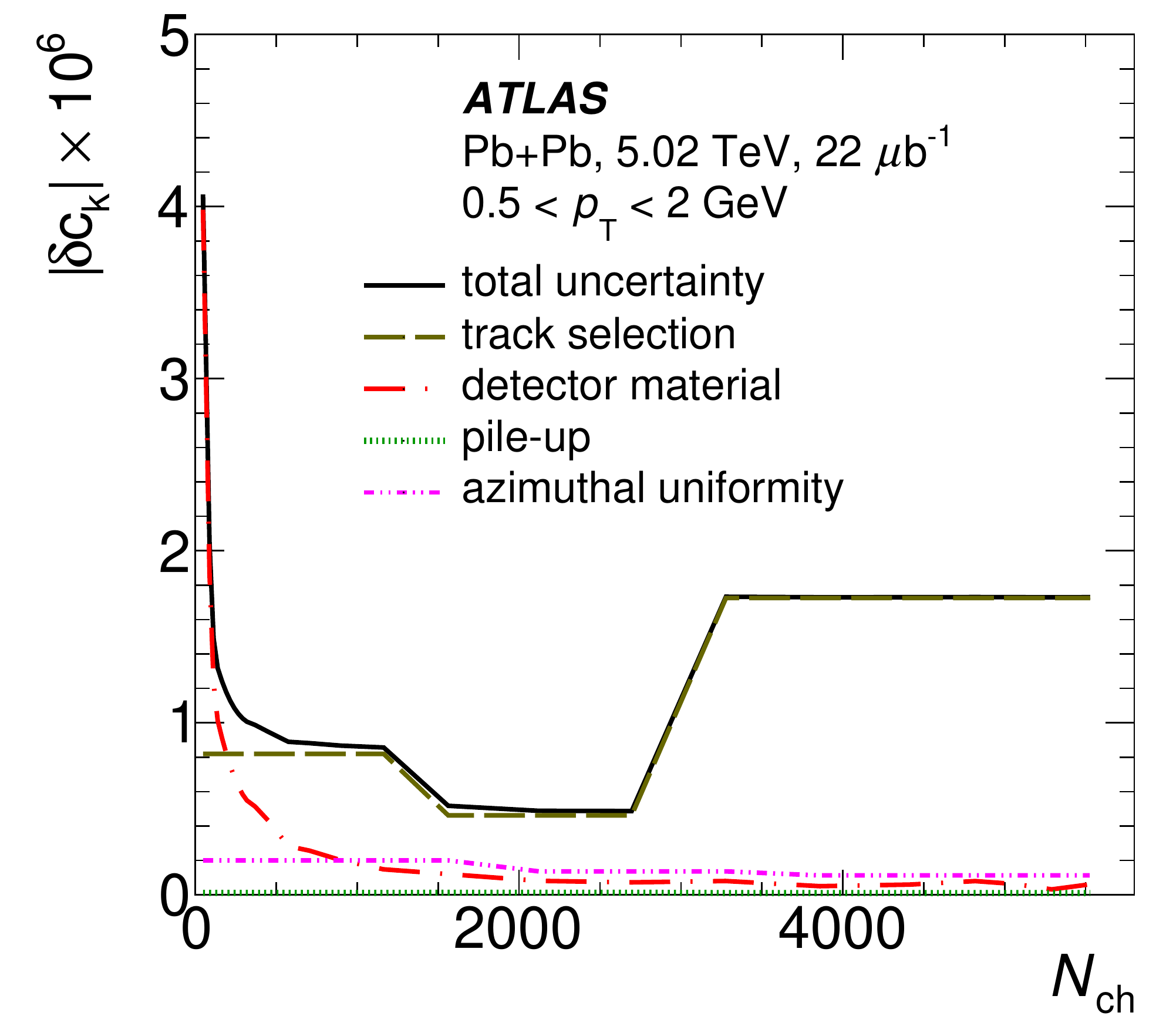} &
\includegraphics[width=0.45\textwidth]{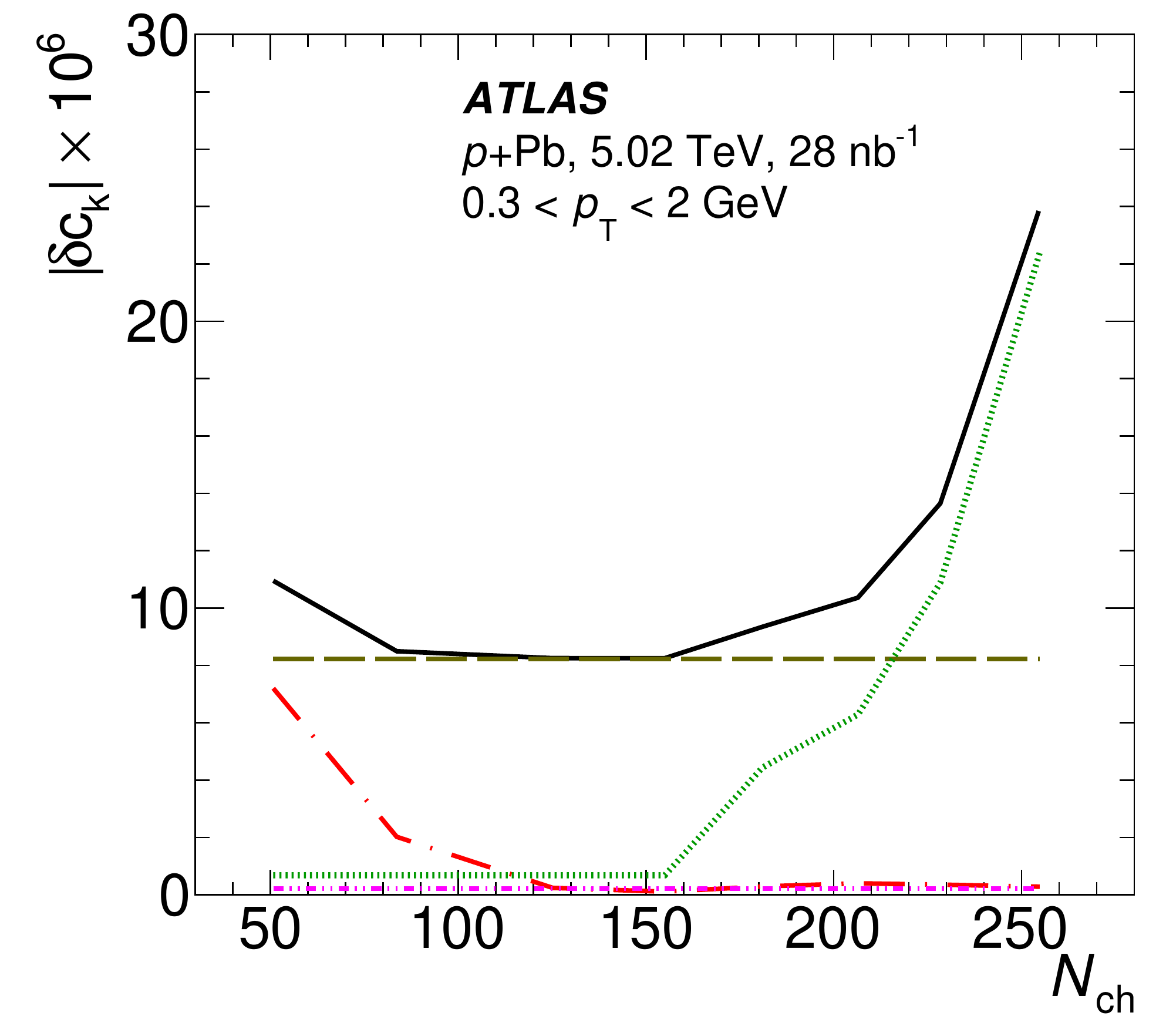}\\
\hspace{1.5cm}(a) &     \hspace{1.5cm}(b) \\
\end{tabular}
\caption{ The systematic uncertainty of \ck\ as a  function of \Nch\ measured with tracks from main \pT\ intervals
in (a) \PbPb\ collisions and in (b) \pPb\ collisions. The total uncertainty is also shown.}
\label{fig:sys_ck}
\end{figure}

\begin{figure}[!htb]
\centering
\begin{tabular}{cc}
 
\includegraphics[width=0.45\textwidth]{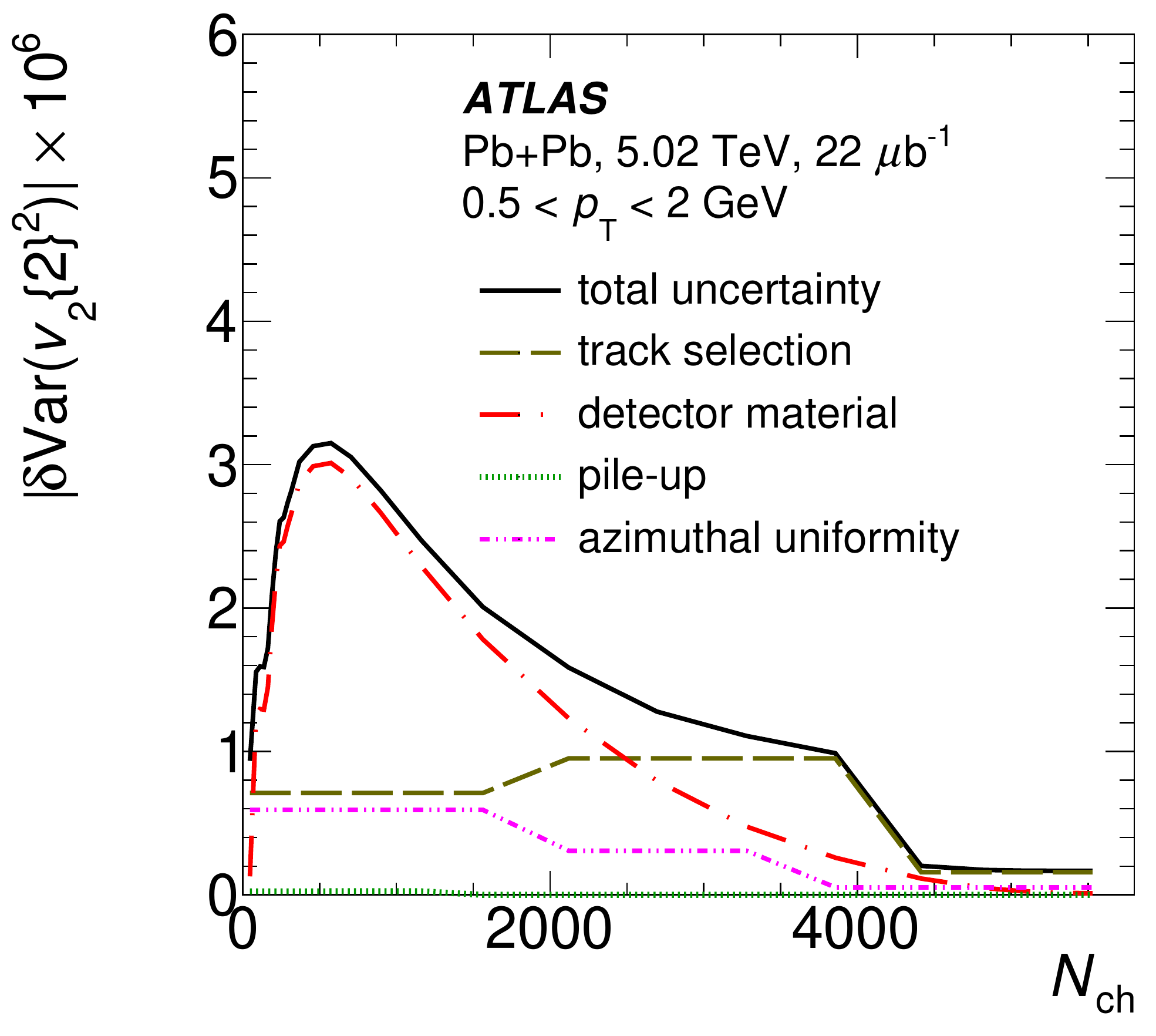} &
\includegraphics[width=0.45\textwidth]{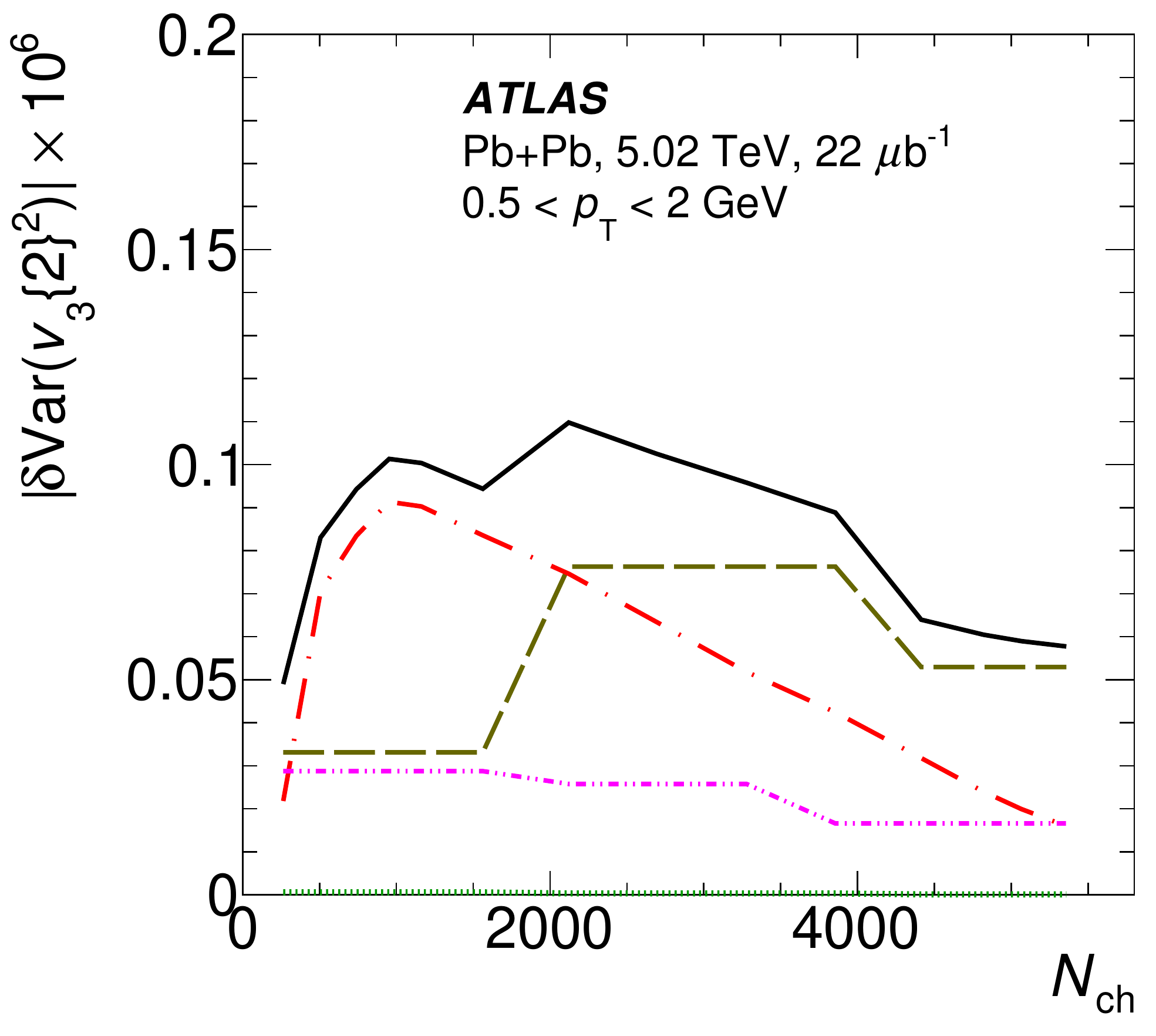}\\
\hspace{1.5cm}(a) &     \hspace{1.5cm}(b) \\
\includegraphics[width=0.45\textwidth]{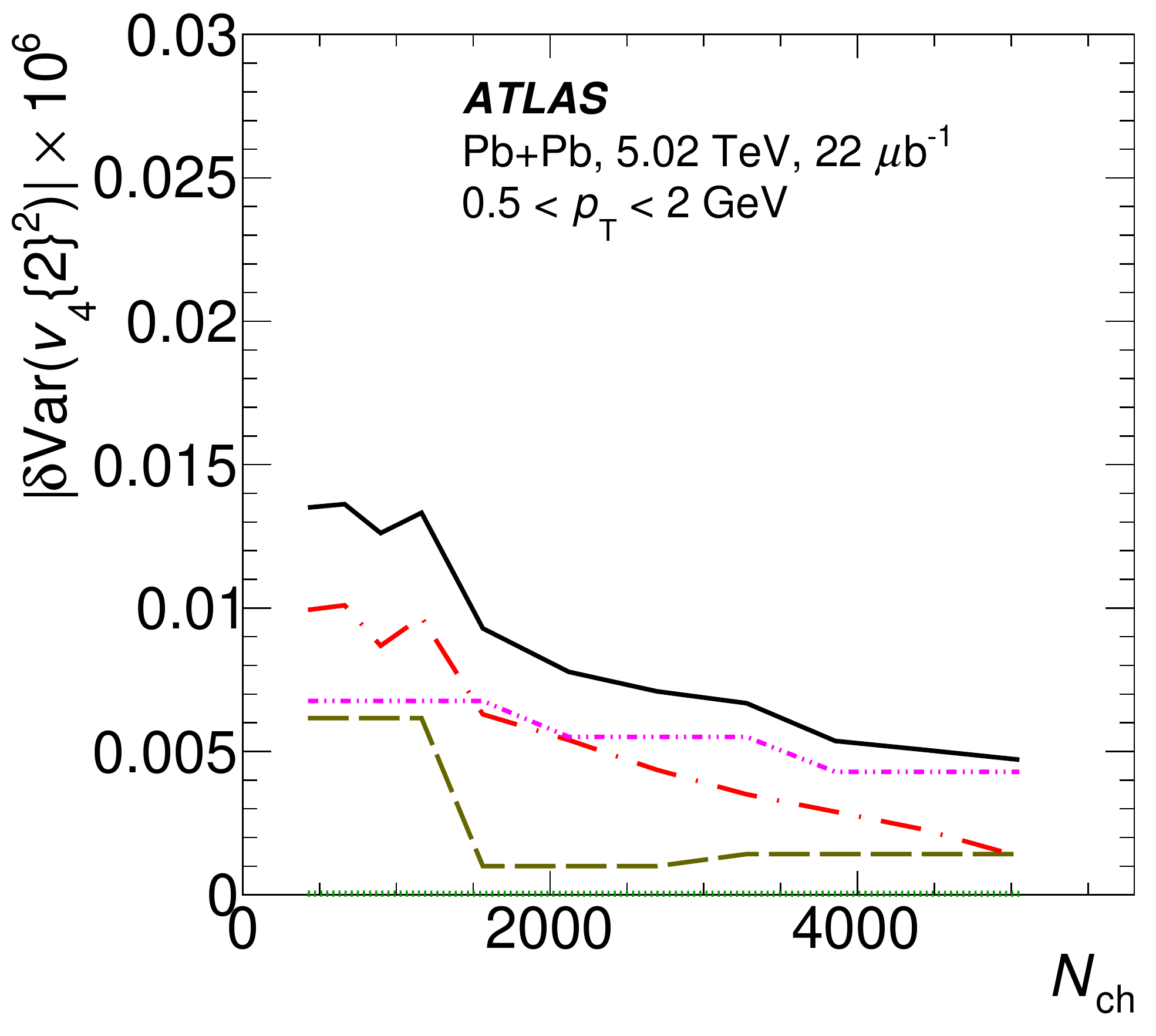}&
\includegraphics[width=0.45\textwidth]{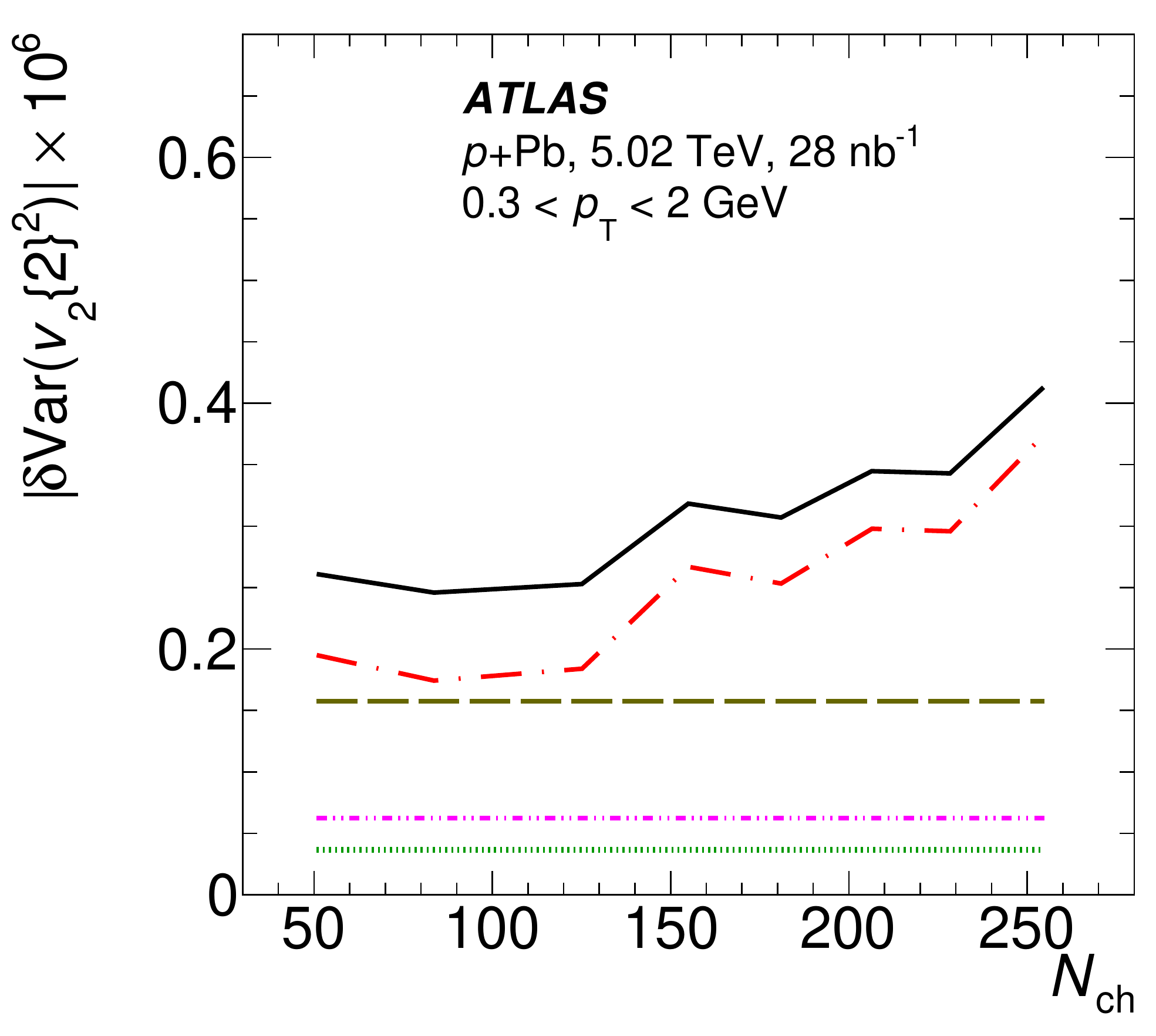}\\
\hspace{1.5cm}(c) &     \hspace{1.5cm}(d) \\
\end{tabular}
\caption{ The systematic uncertainty of \varn{n}\ as a function of \Nch\ measured with tracks from main \pT\ intervals
for each collision system for the (a) second, (b) third, and (c) fourth order harmonics in \PbPb\ collisions,
and for (d) \varn{2} in \pPb\ collisions. The total uncertainty is also shown.}
\label{fig:sys_var}
\end{figure}

\begin{figure}[!htb]
\centering
\begin{tabular}{cc}
\includegraphics[width=0.45\textwidth]{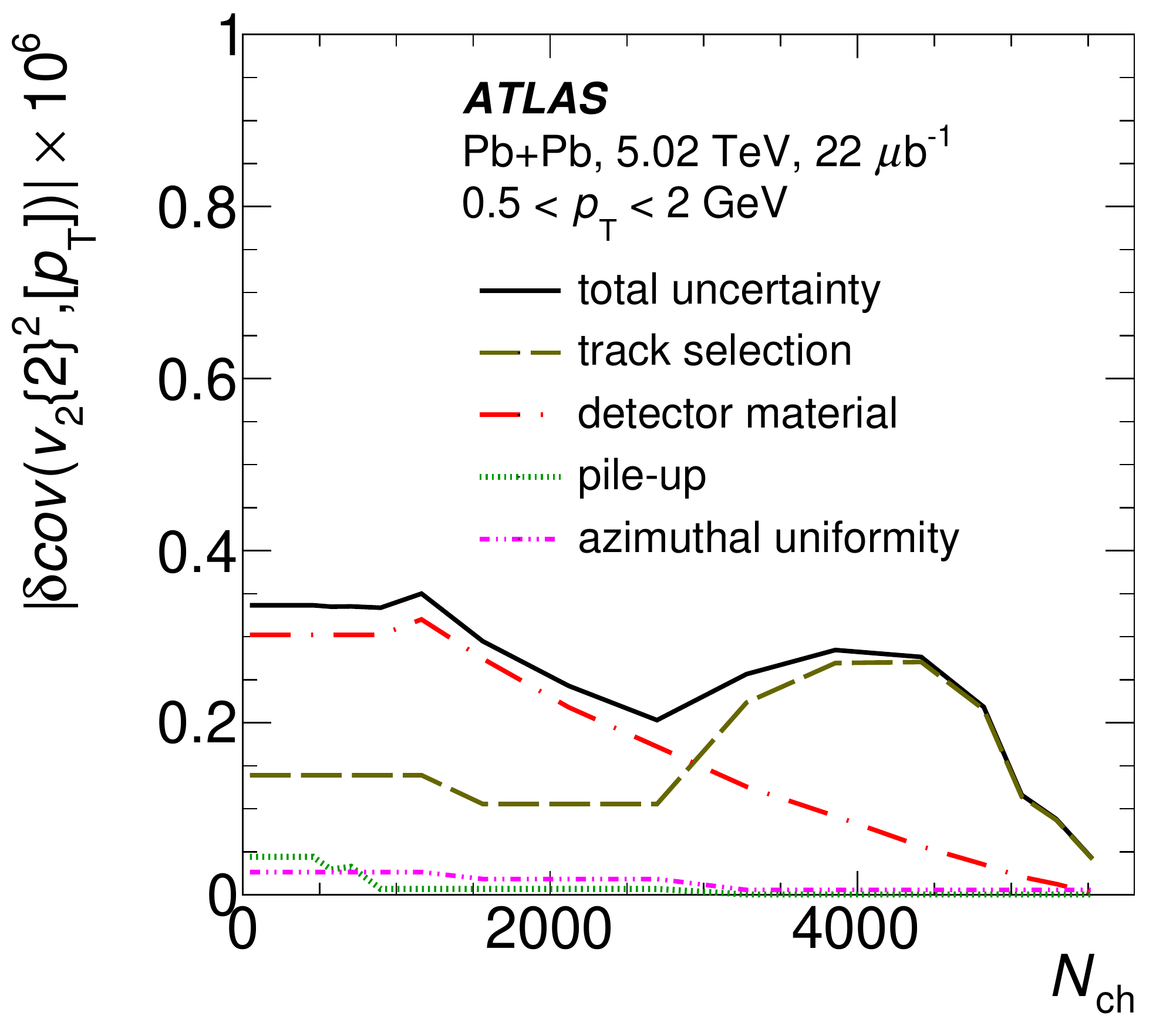} &
\includegraphics[width=0.45\textwidth]{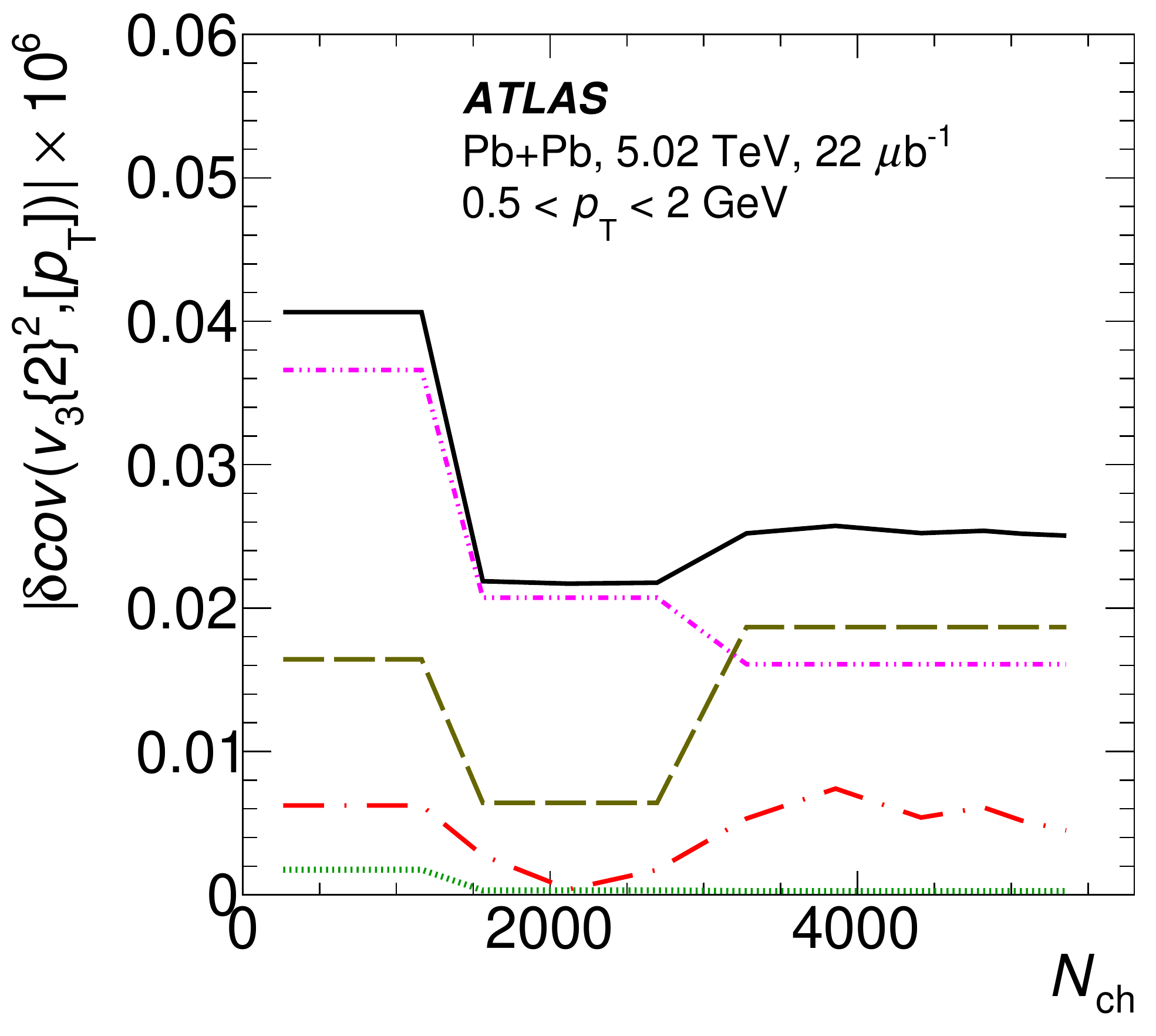}\\
\hspace{1.5cm}(a) &     \hspace{1.5cm}(b) \\
\includegraphics[width=0.45\textwidth]{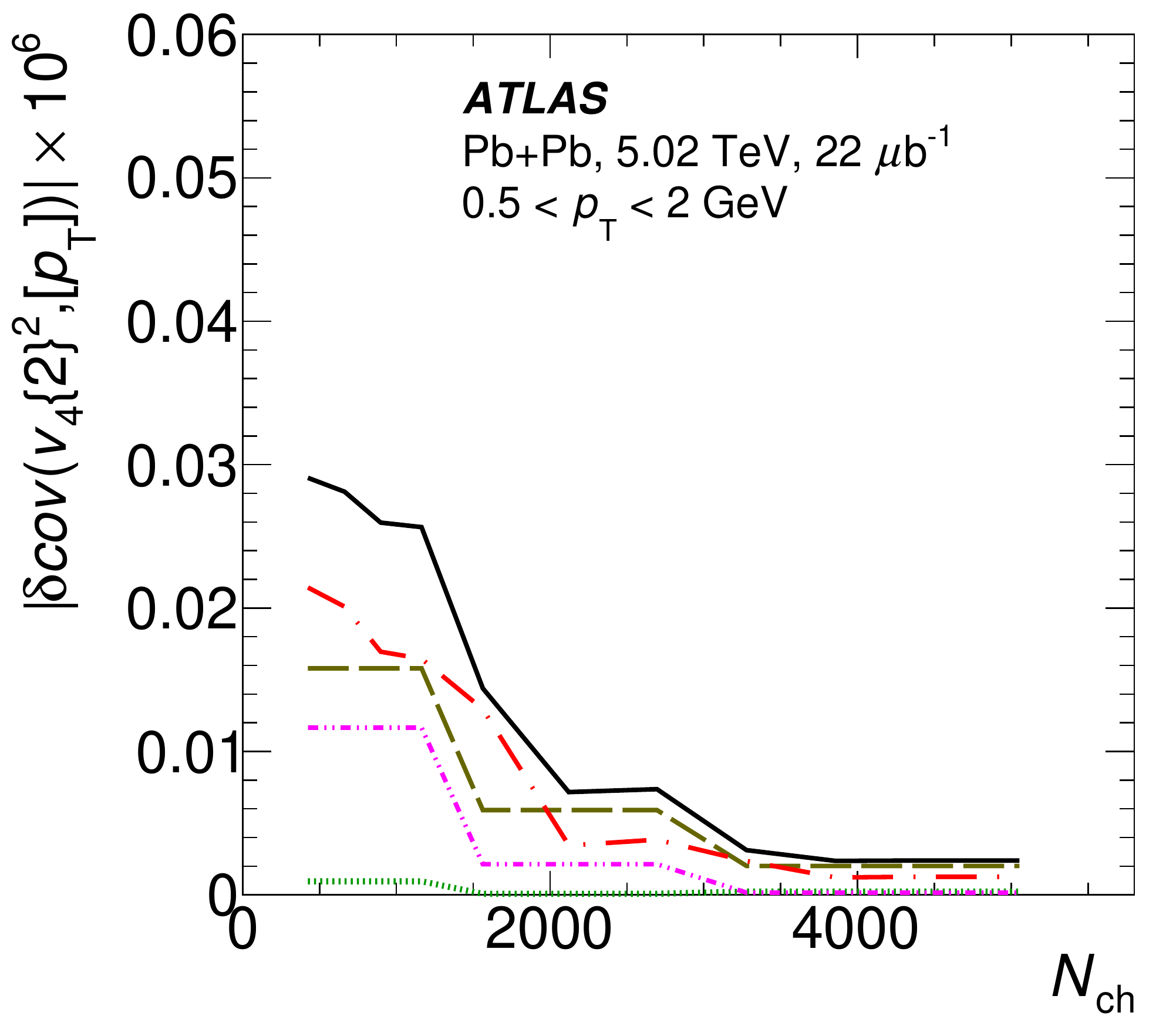}&
\includegraphics[width=0.45\textwidth]{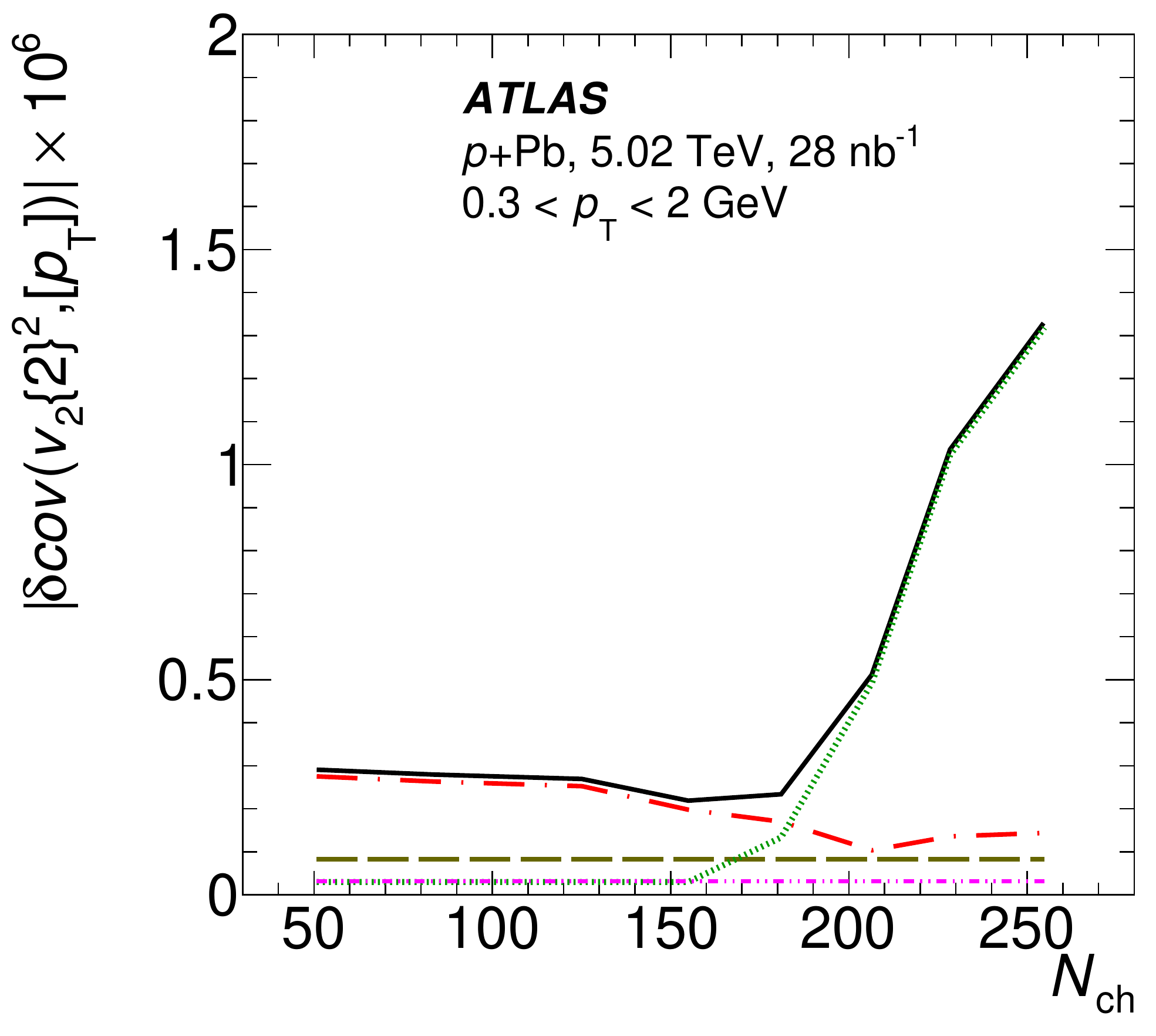}\\
\hspace{1.5cm}(c) &     \hspace{1.5cm}(d) \\
\end{tabular}
\caption{The systematic uncertainty of \covn{n}\ as a function of \Nch\ measured with tracks from main \pT\ intervals
for each collision system for the (a) second, (b) third, and (c) fourth order harmonics in \PbPb\ collisions,
and for (d) \covn{2} in \pPb\ collisions. The total uncertainty is also shown.}
\label{fig:sys_cov}
\end{figure}

This section presents the systematic uncertainties of \ck,
\varn{n}\ and \covn{n}\ for the \PbPb and \pPb collisions
at 5.02 TeV as a function of \Nch.
Each figure shows individual contributions
to the total uncertainty from sources described in Section~\ref{sec:Sys},
i.e. track selection,  detector material, tracking azimuthal non-uniformity
and residual pile-up events.
Figure~\ref{fig:sys_ck} shows contributions to
the systematic uncertainty of \ck\ measured with tracks from the main \pT\ intervals in \PbPb\  and \pPb\ collisions.
The  contributions to the systematic uncertainty of \varn{n}\
as a function of \Nch  for each collision system for the second, third,
and fourth order harmonics in \PbPb\ collisions, and for \varn{2} in \pPb\
collisions are shown in Fig.~\ref{fig:sys_var}.
Figure~\ref{fig:sys_cov}  presents the corresponding systematic
uncertainty of \covn{n}\ for the second, third,
and fourth order harmonics in \PbPb\ collisions, and for \covn{2} in \pPb\ collisions.
 
\clearpage

\printbibliography

\clearpage \input{atlas_authlist}

\end{document}

%% file: atlas_authlist.tex
 
\begin{flushleft}
{\Large The ATLAS Collaboration}

\bigskip

G.~Aad$^\textrm{\scriptsize 101}$,    
B.~Abbott$^\textrm{\scriptsize 128}$,    
D.C.~Abbott$^\textrm{\scriptsize 102}$,    
A.~Abed~Abud$^\textrm{\scriptsize 70a,70b}$,    
K.~Abeling$^\textrm{\scriptsize 53}$,    
D.K.~Abhayasinghe$^\textrm{\scriptsize 93}$,    
S.H.~Abidi$^\textrm{\scriptsize 167}$,    
O.S.~AbouZeid$^\textrm{\scriptsize 40}$,    
N.L.~Abraham$^\textrm{\scriptsize 156}$,    
H.~Abramowicz$^\textrm{\scriptsize 161}$
% ...
\bigskip
\\

% ...
$^{40}$Niels Bohr Institute, University of Copenhagen, Copenhagen; Denmark.\\
% ...
$^{53}$II. Physikalisches Institut, Georg-August-Universit\"{a}t G\"ottingen, G\"ottingen; Germany.\\
% ...
$^{70}$$^{(a)}$INFN Sezione di Pavia;$^{(b)}$Dipartimento di Fisica, Universit\`a di Pavia, Pavia; Italy.\\
% ...
$^{93}$Department of Physics, Royal Holloway University of London, Egham; United Kingdom.\\
% ...
$^{101}$CPPM, Aix-Marseille Universit\'e, CNRS/IN2P3, Marseille; France.\\
$^{102}$Department of Physics, University of Massachusetts, Amherst MA; United States of America.\\
% ...
$^{128}$Homer L. Dodge Department of Physics and Astronomy, University of Oklahoma, Norman OK; United States of America.\\
% ...
$^{156}$Department of Physics and Astronomy, University of Sussex, Brighton; United Kingdom.\\
% ...
$^{161}$Raymond and Beverly Sackler School of Physics and Astronomy, Tel Aviv University, Tel Aviv; Israel.\\
% ...
$^{167}$Department of Physics, University of Toronto, Toronto ON; Canada.\\
% ...
$^{a}$ Also at Borough of Manhattan Community College, City University of New York, New York NY; United States of America.\\
$^{b}$ Also at CERN, Geneva; Switzerland.\\
% ...

\end{flushleft}
